\documentclass[journal]{IEEEtran}
\usepackage{cite,amsmath,amssymb,amsfonts,algorithmic,graphicx,mathtools}
\usepackage{textcomp,multicol,multirow,booktabs,makecell,algorithm,subfigure,cleveref,bm}

\RequirePackage[normalem]{ulem}
\RequirePackage{color}\definecolor{RED}{rgb}{1,0,0}\definecolor{BLUE}{rgb}{0,0,1}
\RequirePackage{settobox}
\RequirePackage{letltxmacro}

\begin{document}

\title{Motor Imagery Classification for Asynchronous EEG-Based Brain-Computer Interfaces}

\author{Huanyu~Wu, Siyang~Li, and Dongrui~Wu
\thanks{H.~Wu, S.~Li and D.~Wu are with the Ministry of Education Key Laboratory of Image Processing and Intelligent Control, School of Artificial Intelligence and Automation, Huazhong University of Science and Technology, Wuhan 430074, China. They are also with Shenzhen Huazhong University of Science and Technology Research Institute, Shenzhen 518063, China.}
\thanks{D.~Wu is the corresponding author (e-mail: drwu09@gmail.com).}}

\maketitle

\begin{abstract}
Motor imagery (MI) based brain-computer interfaces (BCIs) enable the direct control of external devices through the imagined movements of various body parts. Unlike previous systems that used fixed-length EEG trials for MI decoding, asynchronous BCIs aim to detect the user's MI without explicit triggers. They are challenging to implement, because the algorithm needs to first distinguish between resting-states and MI trials, and then classify the MI trials into the correct task, all without any triggers. This paper proposes a sliding window prescreening and classification (SWPC) approach for MI-based asynchronous BCIs, which consists of two modules: a prescreening module to screen MI trials out of the resting-state, and a classification module for MI classification. Both modules are trained with supervised learning followed by self-supervised learning, which refines the feature extractors. Within-subject and cross-subject asynchronous MI classifications on four different EEG datasets validated the effectiveness of SWPC, i.e., it always achieved the highest average classification accuracy, and outperformed the best state-of-the-art baseline on each dataset by about 2\%.
\end{abstract}

\begin{IEEEkeywords}
Brain-computer interface, electroencephalogram, motor imagery, self-supervised learning
\end{IEEEkeywords}

\IEEEpeerreviewmaketitle

\section{Introduction}

\begin{figure}[hptb] \centering
\includegraphics[width=\linewidth,clip]{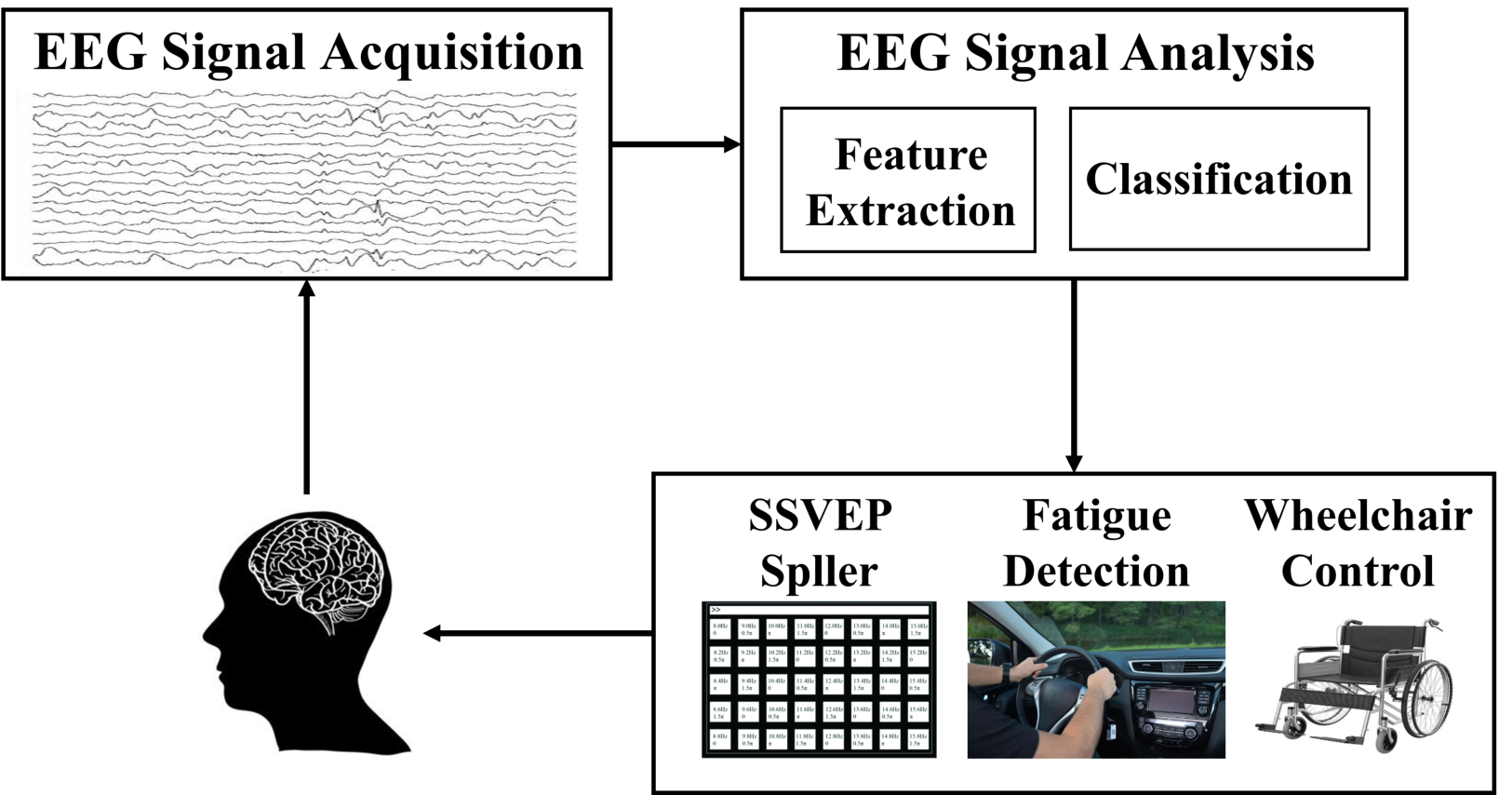}
\caption{Flowchart of a closed-loop EEG-based BCI system. } \label{fig:BCI}
\end{figure}

Non-invasive electroencephalogram (EEG) based brain-computer interfaces (BCIs) have made rapid progress recently~\cite{Lance2012,drwuMITLBCI2022,drwuPIEEE2023}. As shown in Fig.~\ref{fig:BCI}, a closed-loop BCI system usually consists of three components: signal acquisition, signal analysis, and external device control. Signal analysis further includes signal preprocessing, feature extraction, and classification. Both traditional classifiers, e.g., linear discriminant analysis and support vector machine, and deep learning, e.g., EEGNet\cite{Lawhern2018}, have been used.

There are three classical BCI paradigms: motor imagery (MI), event-related potential, and steady-state visual evoked potential. The paper focuses on MI, where the user imagines the movement of various body parts, e.g., left/right hand, both feet or tongue, to elicit different EEG patterns and hence to control external devices. It has been used in upper limb robotic rehabilitation~\cite{Ang2009}, text input~\cite{Zhang2018}, wheelchair control~\cite{Palumbo2021}, etc.

Most existing MI-based BCIs use specific triggers to indicate the start and end of each MI trial, which may be inconvenient in practice. For example, when an MI-based BCI is used to navigate a wheelchair, the control commands should be sent out whenever the user wants, instead of only after some triggers.

\begin{figure}[htpb]   \centering
    \includegraphics[width=.95\linewidth,clip]{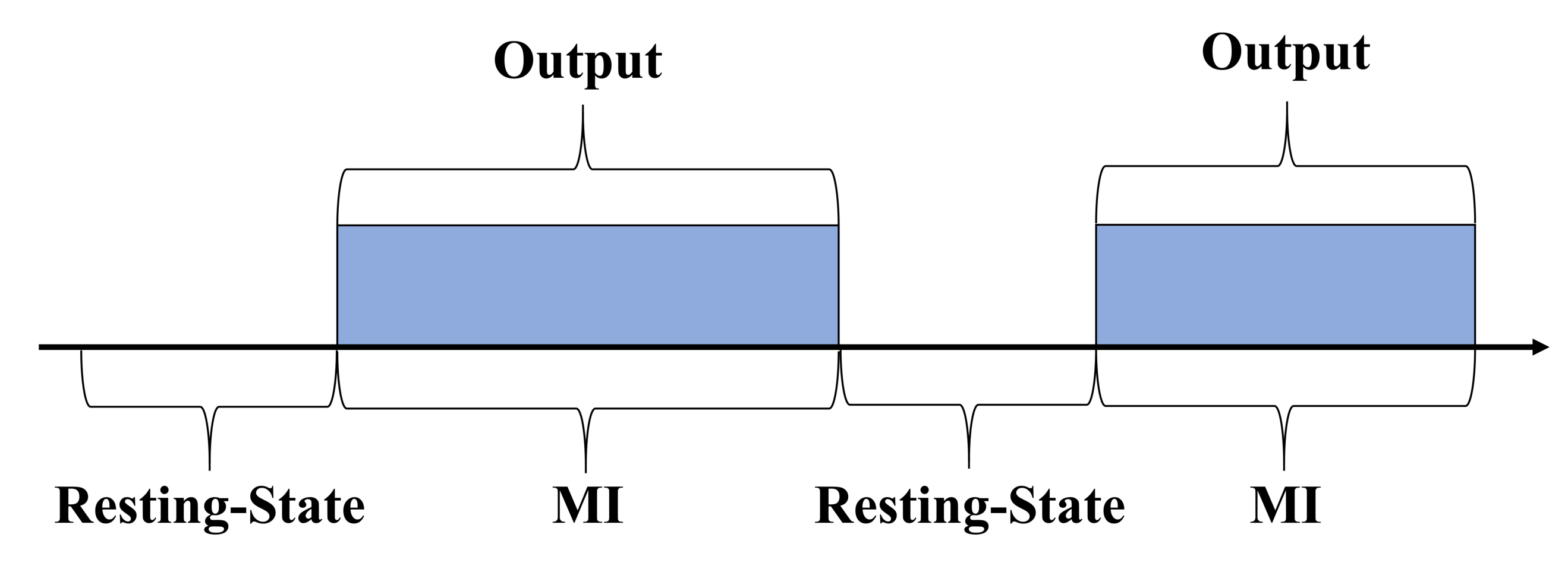}
    \caption{Illustration of asynchronous MI classification. The user may switch between resting-state and MI at any unknown time.}   \label{fig:AS}
\end{figure}

Mason \emph{et al.}~\cite{Mason2000} first proposed the concept of asynchronous BCIs. As shown in Fig.~\ref{fig:AS}, a subject generates control signals by consciously changing his/her mental state: when the subject starts MI, the BCI system detects it and completes the corresponding instruction; otherwise, it keeps still.

Asynchronous MI-based BCIs are challenging to implement, because the algorithm needs to first distinguish between resting-states (no MI) and MI trials, and then classify the MI trials into the correct task, all without any triggers. Very few studies have appeared in the literature. For example, Sugiura \emph{et al.}~\cite{Sugiura2007} adopted a hierarchical hidden Markov model, and Saa and Cetin~\cite{Saa2013} proposed conditional random fields and latent dynamic conditional random fields for EEG classification in asynchronous BCIs.

This paper proposes a sliding window prescreening and classification (SWPC) approach for asynchronous MI-based BCIs, which consists of two modules:
\begin{enumerate}
\item \emph{Prescreening module}, where a classifier with a fixed window length, trained with both supervised learning and self-supervised learning (SSL), is used to prescreen MIs from the resting-state. If the output probability exceeds a threshold, then the EEG trial is sent to the next module for classification.
\item \emph{Classification module}, where a classifier, also trained with supervised learning and SSL, is used for MI classification.
\end{enumerate}
Within-subject and cross-subject experiments on four MI datasets demonstrated the effectiveness of SWPC, particularly SSL, to refine the feature extractors.

The remainder of this paper is organized as follows. Section~\ref{sec:methodology} introduces our proposed SWPC. Section~\ref{sec:experiments} validates the performance of SWPC on four MI datasets. Finally, Section~\ref{sec:conclusion} draws conclusions and points out future research directions.

\section{Methodology} \label{sec:methodology}

This section introduces our proposed SWPC for asynchronous MI-based BCIs. The code is publicly available at https://github.com/why135724/SWPC.

\subsection{Flowchart of SWPC}  \label{subsec:SWPC}

\begin{figure*}[htpb]   \centering
    \includegraphics[width=0.9\linewidth,clip]{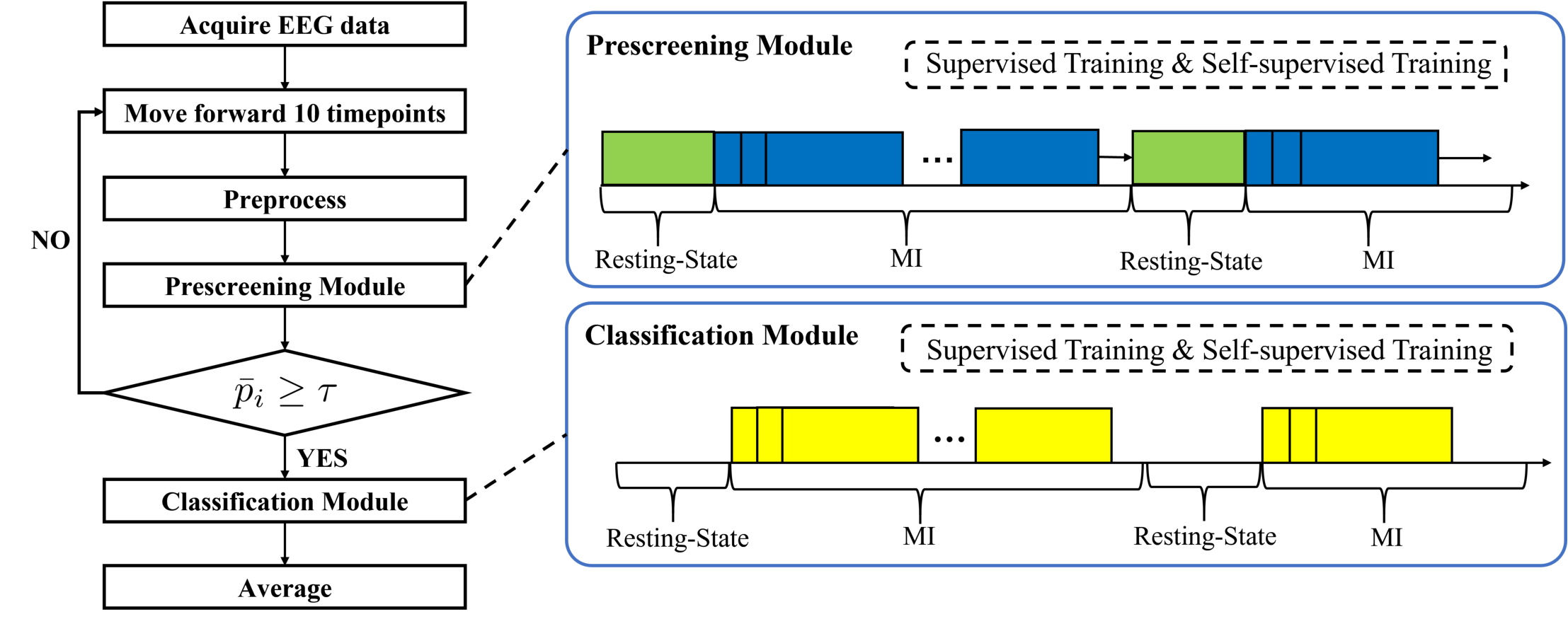}
    \caption{SWPC for asynchronous MI-based BCIs.}   \label{fig:SWPC}
\end{figure*}

Fig.~\ref{fig:SWPC} shows the flowchart of our proposed SWPC for asynchronous MI-based BCIs. It includes two modules:
\begin{enumerate}
\item \emph{Prescreening module}, where an EEGNet\cite{Lawhern2018} classifier with a fixed window length is trained to screen the MIs out of the resting-state EEG trials.
\item \emph{Classification module}, where another EEGNet classifier is trained to classify the prescreened MI trials.
\end{enumerate}

\subsection{Problem Setting} \label{subsec:problem setup}

Two training sets are used in SWPC.

The first training set, $\mathcal{D}^s=\{(X_i^s,y_i^s) \}^{n_s}_{i=1}$, is used in the classification module to classify potential MI trials into different MI tasks. It consists of $n_s$ labeled MI trials $X_i^s \in \mathbb{R}^{ch \times ts}$ and the corresponding labels $y_i^s$, where $ch$ is the number of EEG channels, and $ts$ the number of time domain samples.

The second training set, $\bar{\mathcal{D}}^s=\{(\bar{X}_i^s,\bar{y}_i^s) \}^{2n_s}_{i=1}$, is used in the prescreening module to distinguish MI trials from the resting-state. It consists of $n_s$ labeled MI trials from $\mathcal{D}^s$, and another $n_s$ resting-state trials ${\bar{X}_i^s} \in \mathbb{R}^{ch \times ts}$ adjacent to the MI trials. $\bar{y}_i^{s}\in\{0,1\}$ (0 denotes resting-state, and 1 denotes MI) is the label of $\bar{X}_i^s$.

The test data $X^t \in \mathbb{R}^{ch \times fl}$ is a long EEG data stream with $fl$ time domain samples, where usually $fl\gg ts$. It does not include any triggers, so we do not know when an MI trial starts. The goal is to correctly identify the MI periods and further classify them into specific MI tasks.

\begin{figure}[htpb] \centering
\includegraphics[width=1.0\linewidth,clip]{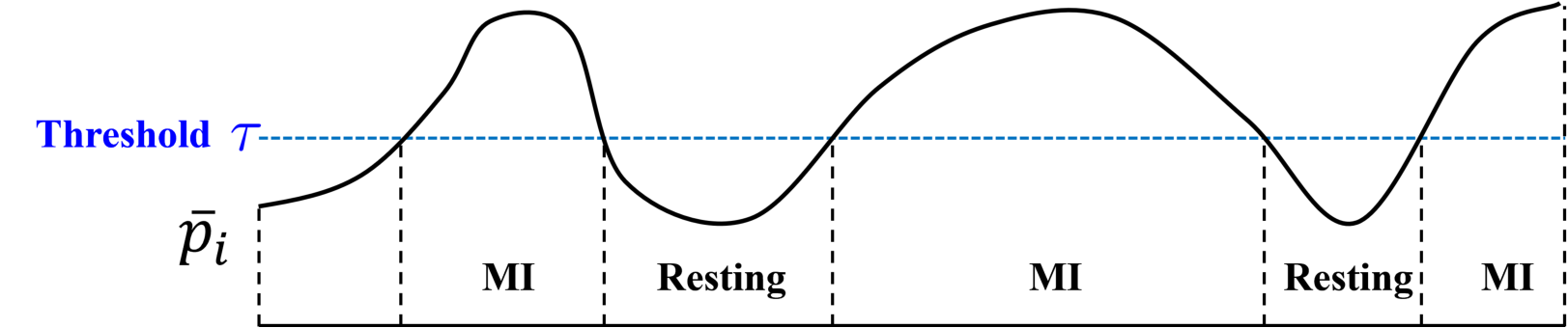}
\caption{Usage of the prescreening probability $\bar{p}_i$. } \label{fig:SWPC_th}
\end{figure}

To simplify the problem, we split $X^t$ with sliding window length $L_{w}$ and step $10$ to get $n_t$ test trials $\mathcal{D}_t=\{ X_i^t \}^{n_t}_{i=1}$. Each trial $X_i^t$ is then passed to the prescreening module, which outputs the probability $\bar{p}_i$ of $X_i^t$ being MI. If $\bar{p}_i$ exceeds a threshold $\tau$, as shown in Fig.~\ref{fig:SWPC_th}, then $X_i^t$ is further passed to the classification module. If multiple successive trials have $\bar{p}_i\ge \tau$, then their corresponding classification probabilities are averaged as the final output. The predicted label for $X_i$ is denoted as $\hat{y}_i^t$.

\subsection{The Prescreening Module}  \label{subsec:SSL for IM}

\begin{figure}[htpb]   \centering
    \includegraphics[width=.8\linewidth,clip]{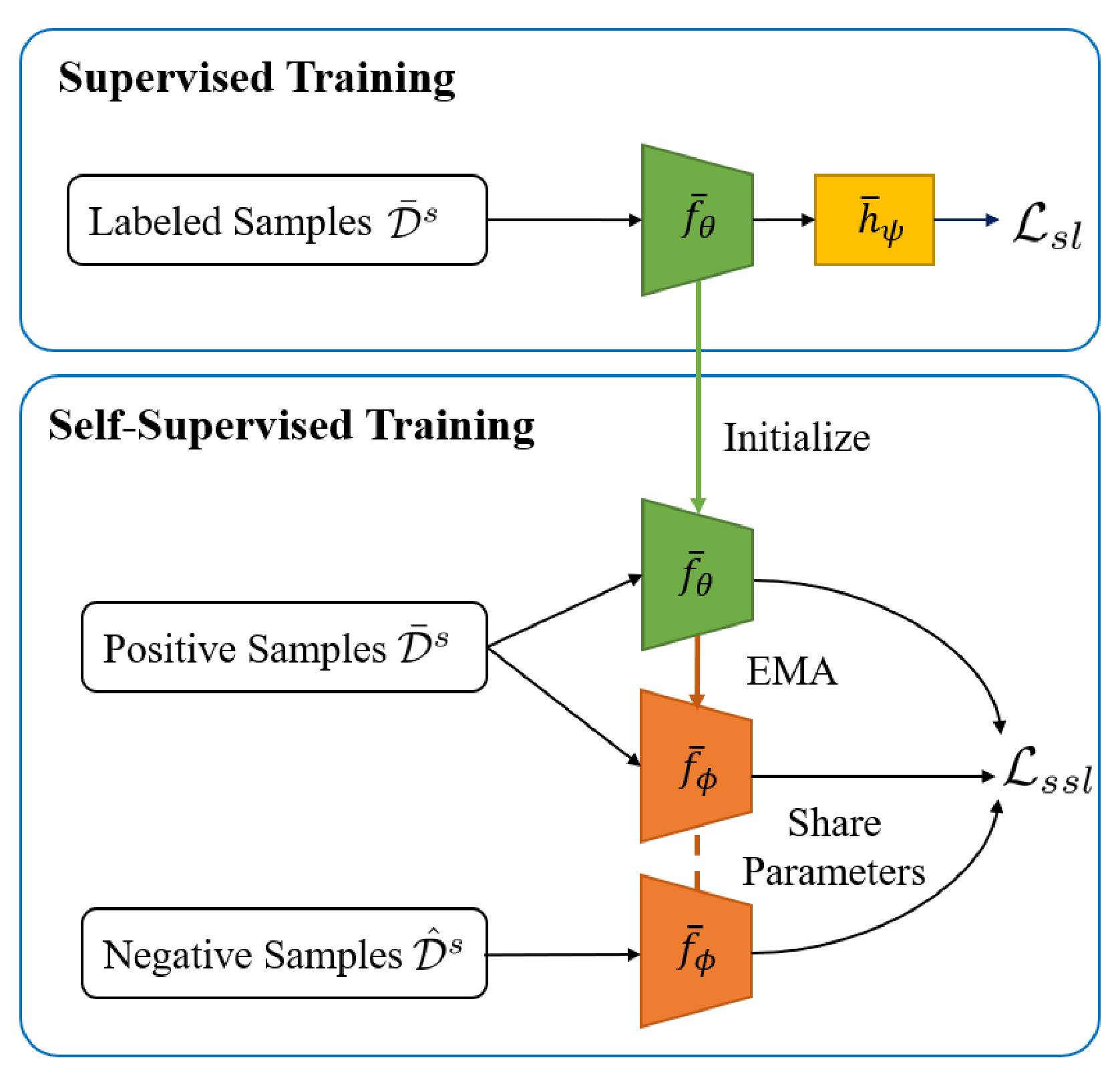}
    \caption{Training of the prescreening module.}   \label{fig:SSL_IM}
\end{figure}

As shown in Fig.~\ref{fig:SSL_IM}, the prescreening module includes first supervised training and then SSL.

Supervised training performs binary classification between MI and resting-states. It trains a feature extractor $\bar{f}_{\boldsymbol{\theta}}$ and a classifier $\bar{h}_{\boldsymbol{\psi}}$ on $\bar{\mathcal{D}}^s$, using the cross-entropy loss:
\begin{align}
\mathcal{L}_{sl}(\bar{h}_{\boldsymbol{\psi}},\bar{f}_{\boldsymbol{\theta}}) = -\sum^{2n_s}_{i=1}\bar{y}_i^s\log_{}{(\bar{h}_{\boldsymbol{\psi}}(\bar{f}_{\boldsymbol{\theta}}(\bar{X}_i^s)))}.  \label{eq:cross entropy loss}
\end{align}
Both $\boldsymbol{\theta}$ and $\boldsymbol{\psi}$ are updated with gradient descent.

SSL is used to fine-tune the feature extractor $\bar{f}_{\boldsymbol{\theta}}$. It first constructs $2n_s$ transition trials (negative samples):
\begin{align}
\hat{X}_i^s = 0.5(\bar{X}_r^s+\bar{X}_m^s), \quad i=1,...,2n_s \label{eq:MCDA}
\end{align}
where $\bar{X}_r^s$ and $\bar{X}_m^s$ are randomly selected resting-state trial and MI trial from $\bar{\mathcal{D}}^s$, respectively. Next, it updates the feature extractor $\bar{f}_{\boldsymbol{\theta}}$ and simultaneously another auxiliary feature extractor $\bar{f}_{\boldsymbol{\phi}}$, on positive samples $\bar{\mathcal{D}}^s$ and negative samples $\hat{\mathcal{D}}^s=\{\hat{X}_i\}_{i=1}^{n_s}$ using the contrastive loss:
\begin{align}
\mathcal{L}_{ssl}(\bar{f}_{\boldsymbol{\theta}},\bar{f}_{\boldsymbol{\phi}}) &= \delta \cdot \exp\left( -\frac{\sum^{2n_s}_{i=1} \vert \bar{f}_{\boldsymbol{\theta}}(\bar{X}_i^s) - \bar{f}_{\boldsymbol{\phi}}(\hat{X}_i^s) \vert^{2}}{2\sigma^{2}} \right) \nonumber\\
&-  \exp\left( -\frac{\sum^{2n_s}_{i=1} \vert \bar{f}_{\boldsymbol{\theta}}(\bar{X}_i^s)-\bar{f}_{\boldsymbol{\phi}}(\bar{X}_i^s)\vert^{2}}{2\sigma^{2}} \right), \label{eq:contrast learning loss}
\end{align}
where $\delta=0.3$ is a hyperparameter controlling the contribution of the negative samples, and $\sigma=2.0$ determines the Gaussian kernel width. Note that $\bar{f}_{\boldsymbol{\theta}}$ and $\bar{f}_{\boldsymbol{\phi}}$ are L2-normalized before entering (\ref{eq:contrast learning loss}), i.e.,
\begin{align}
\bar{f}_{\boldsymbol{\theta}}(\bar{X}_i^s) \leftarrow \frac{\bar{f}_{\boldsymbol{\theta}}(\bar{X}_i^s)}{\Vert \bar{f}_{\boldsymbol{\theta}}(\bar{X}_i^s) \Vert_2}, \quad \bar{f}_{\boldsymbol{\phi}}(\hat{X}_i^s) \leftarrow \frac{\bar{f}_{\boldsymbol{\phi}}(\bar{X}_i^s)}{\Vert \bar{f}_{\boldsymbol{\phi}}(\bar{X}_i^s) \Vert_2}.
\label{eq:L2 norm}
\end{align}

$\bar{f}_{\boldsymbol{\theta}}$ in (\ref{eq:contrast learning loss}) is optimized by gradient descent, whereas $\bar{f}_{\boldsymbol{\phi}}$ is optimized through exponential moving average (EMA):
\begin{align}
\boldsymbol{\phi}^{n+1} \leftarrow  \lambda \cdot \boldsymbol{\phi}^{n} + (1-\lambda) \cdot \boldsymbol{\theta}^{n},  \label{eq:gradient descent target}
\end{align}
where $\lambda = 0.9995$.

For a test EEG trial $X_i^t$, only $\bar{f}_{\boldsymbol{\theta}}$ and $\bar{h}_{\boldsymbol{\psi}}$ are used to compute the prescreening probability $\bar{p}_i$, i.e.,
\begin{align}
\bar{p}_i=\bar{h}_{\boldsymbol{\psi}}(\bar{f}_{\boldsymbol{\theta}}(X_i^t)).
\end{align}

If $\bar{p}_i$ exceeds a threshold $\tau$, then the EEG trial is further passed to the classification module.

\subsection{The Classification Module}  \label{subsec:SSL for CM}

\begin{figure}[htpb]   \centering
    \includegraphics[width=.9\linewidth,clip]{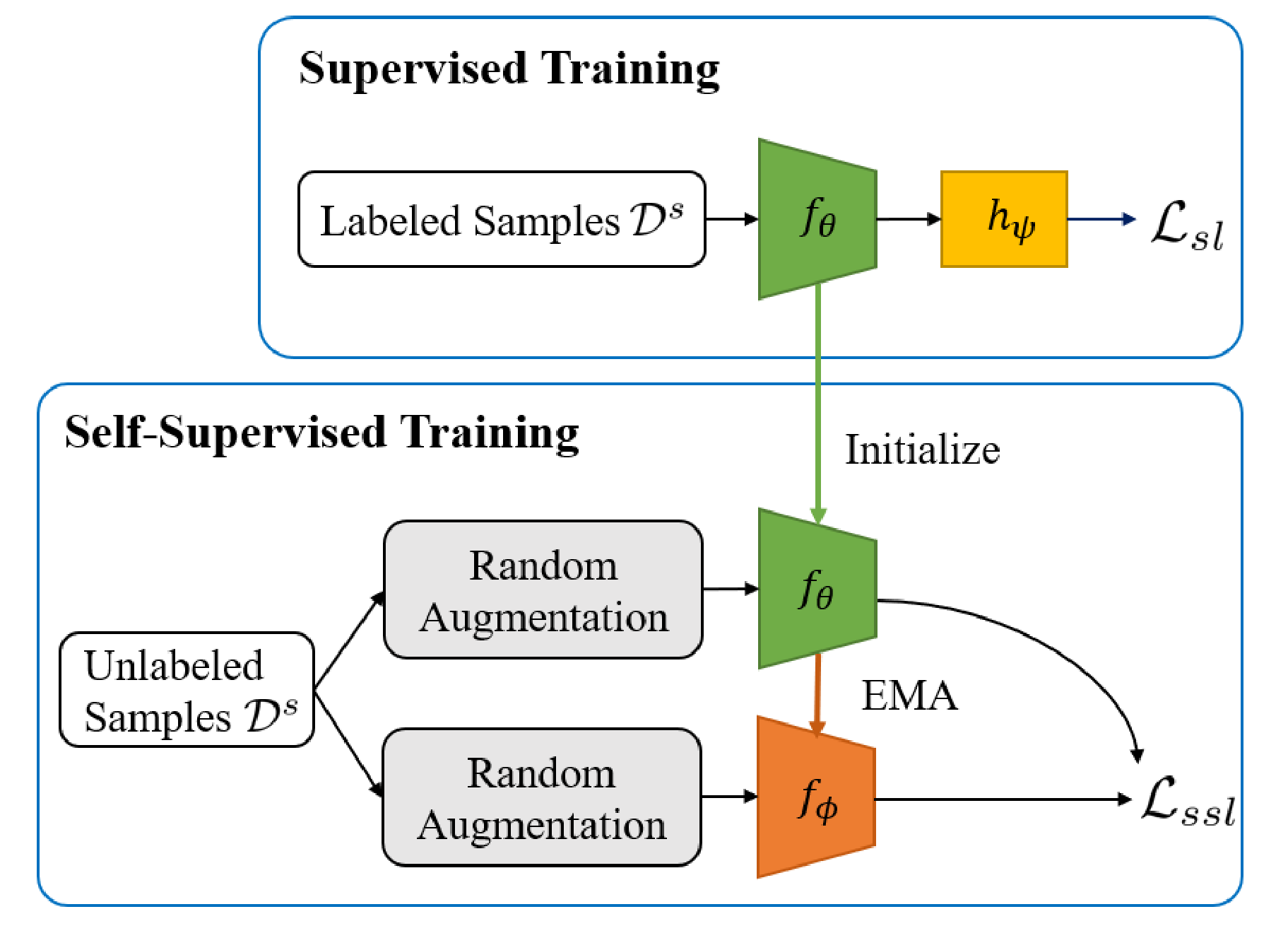}
    \caption{Training of the classification module.}   \label{fig:SSL_CM}
\end{figure}

As shown in Fig.~\ref{fig:SSL_CM}, the training process of the classification module is similar to that of the prescreening module. It also consists of two steps: supervised training and SSL.

Supervised training of the classification module remains the same as supervised training of the prescreening module, except that the EEG trials are classified into different MI tasks, instead of MI and resting-state.

SSL is again used to refine the feature extractor $f_{\boldsymbol{\theta}}$. We replace the construction of negative samples in the prescreening module with data augmentation in the classification module. The following data augmentations are used in this paper:
\begin{enumerate}
    \item Adding noise: Uniform noise $0.5U(-\delta,\delta)$ is added to each element of the feature vector, where $\delta$ is the standard deviation of the original feature.
    \item Scaling amplitude: Each feature is scaled by $0.75$ or $1.25$.
    \item Masking channels: Randomly set all signals in some EEG channels to 0.
    \item Masking segments: Randomly set some segments of the EEG signal to 0.
\end{enumerate}

For each trial $X_{i}^s$, we randomly select two different data augmentations to get $X_{i,1}^s$ and $X_{i,2}^s$. $\{X_{i,1}^s\}^{n_s}_{i=1}$ and $\{X_{i,2}^s\}^{n_s}_{i=1}$ are then L2-normalized using (\ref{eq:L2 norm}) before computing the following contrastive loss:
\begin{equation}
\begin{aligned}
\mathcal{L}_{ssl}(f_{\boldsymbol{\theta}},f_{\boldsymbol{\phi}}) &= - \exp\left( -\frac{\sum^{n_s}_{i=1} \vert f_{\boldsymbol{\theta}}(X_{i,1}^s)-f_{\boldsymbol{\phi}}(X_{i,2}^s)\vert^{2}}{2\sigma^{2}} \right).
\label{eq:contrast learning loss CM}
\end{aligned}
\end{equation}

For an input test trial $X_i^t$, the instantaneous classification probability is
\begin{align}
p_i=h_{\boldsymbol{\psi}}(f_{\boldsymbol{\theta}}(X_i^t)).
\end{align}

To stabilize the output, we average all successive $p_i$ whose corresponding $\bar{p}_i$ exceed $\tau$, i.e.,
\begin{align}
\hat{p}_i = \frac{1}{i-i_0}\sum\limits_{j=i_0}^i p_j, \label{eq:TMA}
\end{align}
where $i_0$ is the smallest index that ensures all $\{\bar{p}_j\}_{j=i_0}^i$ exceed the threshold $\tau$. $\hat{p}_i$ is the final prediction probability for $X_i^t$.

\begin{figure}[htpb]   \centering
    \includegraphics[width=\linewidth,clip]{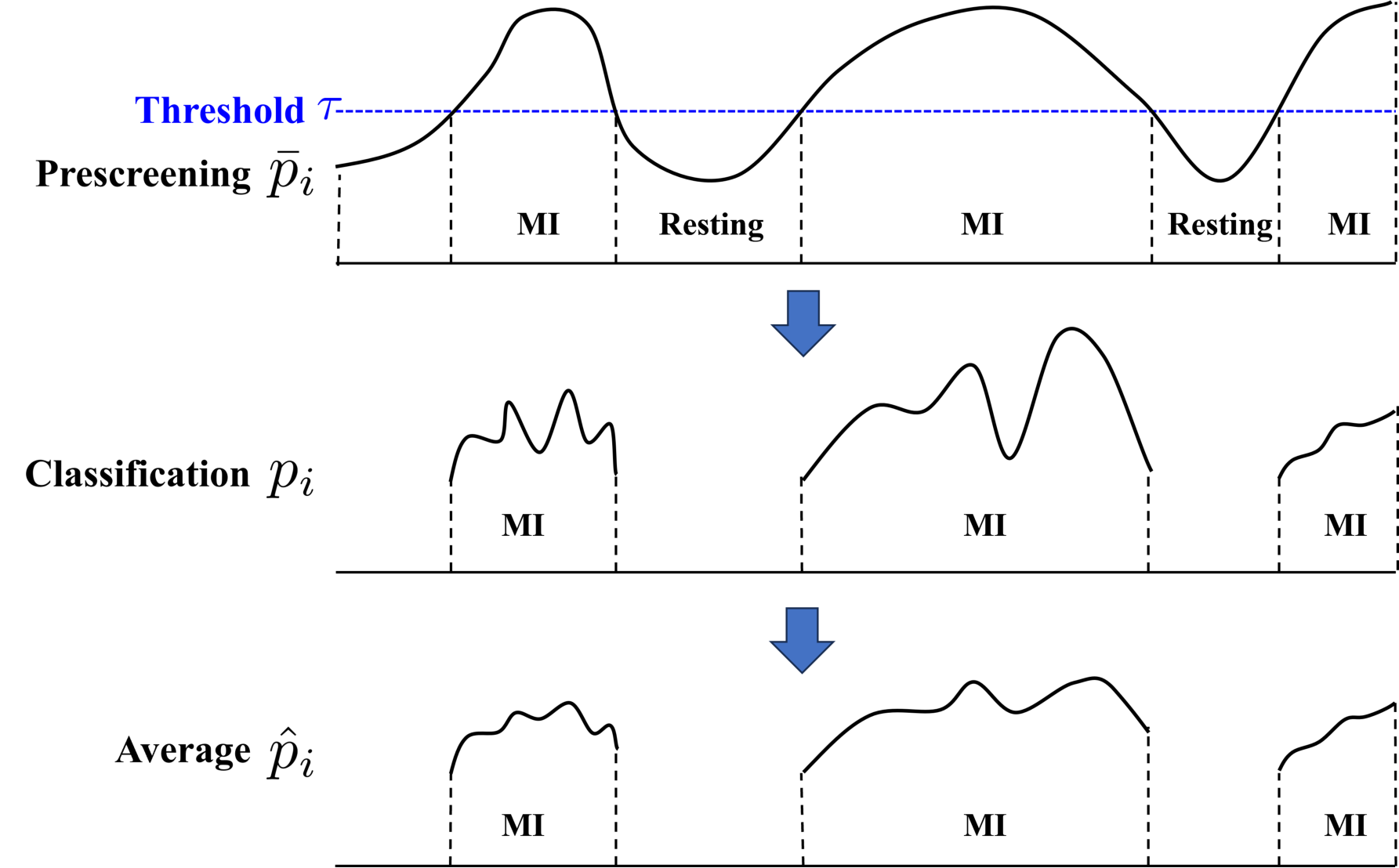}
    \caption{Illustration of computing $\hat{p}_i$.}   \label{fig:TMA}
\end{figure}

More specifically, as shown in Fig.~\ref{fig:TMA}, the process of computing $\hat{p}_i$ is: We pass $X_i^t\in\mathcal{D}_t$ to $\bar{f}_{\boldsymbol{\theta}}$ and $\bar{h}_{\boldsymbol{\psi}}$ to get $\bar{p}_i$. If $\bar{p}_i<\tau$, then we classify the corresponding $X_i^t$ as resting-state; otherwise, we further pass $X_i^t$ to $f_{\boldsymbol{\theta}}$ and $h_{\boldsymbol{\psi}}$ to get $p_i$. $\hat{p}_i$ is then averaged by (\ref{eq:TMA}) and used to derive $\hat{y}_i^t$.

The pseudo-code of SWPC is given in Algorithm~\ref{alg:SWPC}.

\begin{algorithm}[htbp]
    \caption{SWPC for asynchronous MI-based BCIs.}     \label{alg:SWPC}
    \begin{algorithmic}[1]
    \REQUIRE Training set $\mathcal{D}^s=\{ ( X_i^s, y_i^s) \}^{n_s}_{i=1}$;\\
    Training set $\mathcal{\bar{D}}_s=\{ ( \bar{X}_i^s, \bar{y}_i^s) \}^{2n_s}_{i=1}$;\\
    Maximum number of epochs, $\max_{1}$ and $\max_{2}$;\\
	\ENSURE The classification $\{ \hat{y}_i^t \}^{n_t}_{i=1}$.\\
	\STATE \texttt{// Supervised Training of the Prescreening Module}
    \STATE Randomly initialize the feature extractor $\bar{f}_{\boldsymbol{\theta}}$ and the classifier $\bar{h}_{\boldsymbol{\psi}}$;
    \FOR {$k=1,...,\max_{1}$}
    \STATE Pass $\mathcal{\bar{D}}_s$ to $\bar{f}_{\boldsymbol{\theta}}$ and $\bar{h}_{\boldsymbol{\psi}}$ to compute $\mathcal{L}_{sl}(\bar{h}_{\boldsymbol{\psi}},\bar{f}_{\boldsymbol{\theta}})$ in (\ref{eq:cross entropy loss});
    \STATE Update $\boldsymbol{\theta}$ and $\boldsymbol{\psi}$;
    \ENDFOR \\
    \STATE \texttt{// Self-supervised Training of the Prescreening Module}
    \STATE Initialize $\boldsymbol{\phi}$ to $\boldsymbol{\theta}$;
    \STATE Generate ${{\hat X}^s}$ using (\ref{eq:MCDA});
    \FOR {$k=1,...,\max_{2}$}
    \STATE Pass $\{X_i^s\}^{2n_s}_{i=1}$ and $\{\hat{X}_i^s\}^{2n_s}_{i=1}$ to $\bar{f}_{\boldsymbol{\theta}}$ and $\bar{f}_{\boldsymbol{\phi}}$ to compute $\mathcal{L}_{ssl}(\bar{f}_{\boldsymbol{\theta}},\bar{f}_{\boldsymbol{\phi}})$ in (\ref{eq:contrast learning loss});
    \STATE Update $\boldsymbol{\theta}$, and $\boldsymbol{\phi}$ using (\ref{eq:gradient descent target});
    \ENDFOR
    \STATE \texttt{// Supervised Training of the Classification Module}
    \STATE Randomly initialize the feature extractor $f_{\boldsymbol{\theta}}$ and the classifier $h_{\boldsymbol{\psi}}$;
    \FOR {$k=1,...,\max_{1}$}
    \STATE Pass $\mathcal{D}^s$ to $f_{\boldsymbol{\theta}}$ and $h_{\boldsymbol{\psi}}$ to compute $\mathcal{L}_{sl}(h_{\boldsymbol{\psi}},f_{\boldsymbol{\theta}})$ in (\ref{eq:cross entropy loss});
    \STATE Update $\boldsymbol{\theta}$ and $\boldsymbol{\psi}$;
    \ENDFOR \\
    \STATE \texttt{// Self-supervised Training of the Classification Module}
    \STATE Initialize $\boldsymbol{\phi}$ to $\boldsymbol{\theta}$;
    \FOR {$k=1,...,\max_{2}$}
    \STATE Augment $\{X_i^s\}^{n_s}_{i=1}$ twice to get $\{X_{i,1}^s\}^{n_s}_{i=1}$ and $\{X_{i,2}^s\}^{n_s}_{i=1}$;
    \STATE Pass $\{X_{i,1}^s\}^{n_s}_{i=1}$ to $f_{\boldsymbol{\theta}}$ and $\{X_{i,2}^s\}^{n_s}_{i=1}$ to $f_{\boldsymbol{\phi}}$, and compute $\mathcal{L}_{ssl}(f_{\boldsymbol{\theta}},f_{\boldsymbol{\phi}})$ in (\ref{eq:contrast learning loss CM});
    \STATE Update $\boldsymbol{\theta}$, and $\boldsymbol{\phi}$ using (\ref{eq:gradient descent target});
    \ENDFOR
    \STATE \texttt{// Test Sample Classification}
    \STATE Split $X^t$ with sliding window length $L_{w}$ and step size $10$ to get $\mathcal{D}_t=\{ X_i^t \}^{n_t}_{i=1}$;
    \STATE Set $i_0=0$;
    \FOR {$i=1,...,n_t$}
    \STATE Pass $X_i^t$ to $\bar{f}_{\boldsymbol{\theta}}$ and $\bar{h}_{\boldsymbol{\psi}}$ to get $\bar{p}_i$;
    \IF {$\bar{p}_i \geq \tau$}
    \IF {$i_0==0$}
    \STATE {Set $i_0=i$;}
    \ENDIF
    \STATE {Pass $X_i^t$ to $f_{\boldsymbol{\theta}}$ and $h_{\boldsymbol{\psi}}$ to compute $p_i$;}
    \STATE {Compute $\hat{p}_i$ using (\ref{eq:TMA});}
    \STATE {Obtain $\hat{y}_i^t$ from $\hat{p}_i$;}
    \ELSE
    \STATE{Set $\hat{y}_i^t$ as resting-state;}
    \STATE{Set $i_0=0$;}
    \ENDIF
    \ENDFOR
    \end{algorithmic}
\end{algorithm}

\section{Experiments}   \label{sec:experiments}

This section evaluates the performance of our proposed SWPC on four public MI datasets in both within-subject and cross-subject classification.

\subsection{Datasets} \label{subsec:datasets}

Four public datasets from BNCI-Horizon\footnote{http://www.bnci-horizon-2020.eu/database/data-sets}, summarized in Table~\ref{tab:dataset}, were used in our experiments:
\begin{enumerate}
\item MI1 was the 001-2014 dataset~\cite{Tangermann2012} recorded from 9 subjects. Each session included 6 runs separated by short breaks. EEG signals were sampled at 250Hz. Only two classes (left-hand and right-hand) were used.
\item MI2 was also the 001-2014 dataset, but with all four classes, i.e., left-hand, right-hand, feet, and tongue.
\item MI3 was the 002-2014 dataset~\cite{Steyrl2016a} recorded from 14 subjects. Each session included 8 runs separated by short breaks. EEG signals were sampled at 512Hz. Only two classes (right-hand and both feet) were used. Subject 10 was removed, as his/her results were close to random.
\item MI4 was the 004-2014 dataset~\cite{Leeb2007a} recorded from 9 subjects. EEG signals were sampled at 250Hz. Only two classes (left-hand and right-hand) were used. We deleted the results of the 2nd and 3rd subjects for the same reason.
\end{enumerate}
Note that Subjects 5 and 6 in MI1, Subject 10 in MI3, and Subjects 2 and 3 in MI4, were removed because their results were close to random.

All EEG signals were bandpass-filtered between 8Hz and 30Hz, and then notch-filtered at 50Hz.

\begin{table}[htpb] \centering
\caption{Summary of the four MI datasets.} \label{tab:dataset}
\begin{tabular}{c|c|c|c|c} \Xhline{1.2pt}
Dataset & \begin{tabular}[c]{@{}c@{}}Number of \\ Subjects\end{tabular} & \begin{tabular}[c]{@{}c@{}}Number of \\ Channels\end{tabular} & \begin{tabular}[c]{@{}c@{}}Trials per \\ Subject\end{tabular} & \begin{tabular}[c]{@{}c@{}}Number of \\ Classes\end{tabular} \\ \hline
MI1      & 9                & 22               & 144                & 2               \\
MI2      & 9                & 22               & 288                & 4               \\
MI3      & 14               & 15               & 100                & 2               \\
MI4      & 9               & 3               & 60                & 2               \\ \Xhline{1.2pt}
\end{tabular}
\end{table}

\subsection{Performance Evaluation} \label{subsec:Performance Evaluation}

\begin{figure}[htpb]   \centering
    \includegraphics[width=\linewidth,clip]{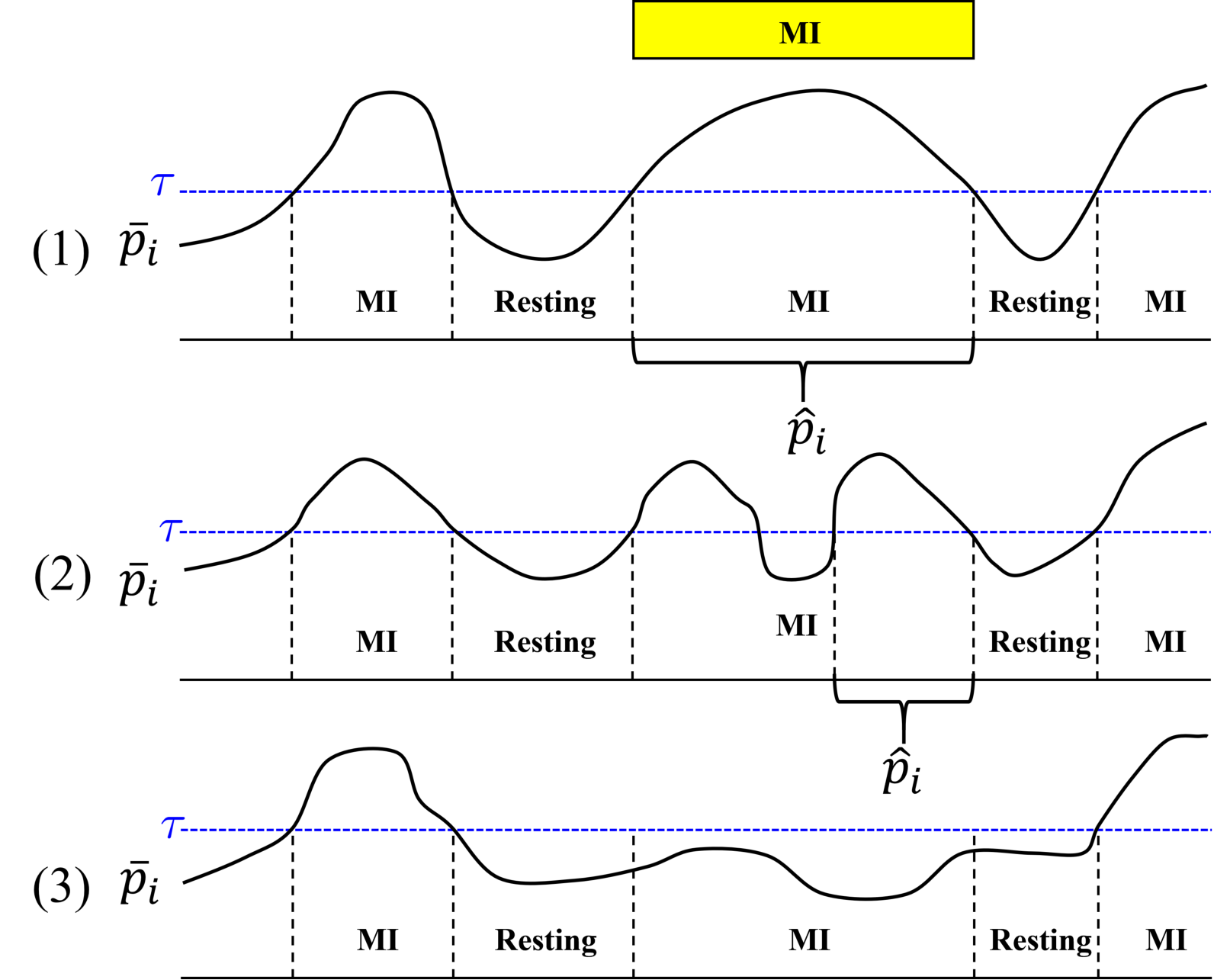}
    \caption{Illustration of computing the classification accuracy in testing.}   \label{fig:ACCcase}
\end{figure}

The classification accuracy (ACC) was used as the performance metric. The specific computation details are illustrated in Fig.~\ref{fig:ACCcase}, where the yellow bar indicates the true MI period:
\begin{enumerate}
    \item When all sliding windows during the true MI period have $\bar{p}_i\ge\tau$, $\hat{p}_i$ corresponding to the last sliding window with $\bar{p}_i\ge\tau$ is used to evaluate the classification accuracy.
    \item When the true MI period is broken into two or more intervals, during each of which all $\bar{p}_i\ge\tau$, $\hat{p}_i$ corresponding to the last sliding window with $\bar{p}_i\ge\tau$ in the last interval is used to evaluate the classification accuracy.
    \item When no sliding window in the MI period has $\bar{p}_i\ge\tau$, the classification is `wrong'.
\end{enumerate}

\subsection{Algorithms} \label{subsec:baseline}

SWPC used the EEGNet backbone. Supervised learning used learning rate 0.0005 and early stopping with patience 30. SSL used learning rate 0.00005 and 40 training epochs.

SWPC was compared with the following 11 approaches:
\begin{enumerate}
    \item Continuous EEG classification (CEC)~\cite{Townsend2004}, which uses CSP and thresholding to identify MIs in EEGs.
    \item Joint training scheme (JTS)~\cite{Cheng2018}, which combines the transitional imagery data (between the resting state and MI) with the resting state to train a binary classifier.
    \item Bootstrap Your Own Latent (BYOL)\footnote{https://github.com/deepmind/deepmind-research/tree/master/byol}~\cite{Grill2020}.
    \item Simple framework for Contrastive Learning of Representations (SimCLR)\footnote{https://github.com/google-research/simclr}~\cite{Chen2020}.
    \item Momentum Contrast (MoCo)\footnote{https://github.com/facebookresearch/moco}~\cite{He2020}.
    \item ContraWR\footnote{https://github.com/ycq091044/ContraWR}~\cite{Yang2023}.
    \item Self-supervised contrastive learning (SSCL)~\cite{Lotey2022}, which was proposed for cross-session MI classification.
    \item Ou2022~\cite{Ou2022}, an SSL approach for MI classification.
    \item Model-agnostic meta-learning (MAML)~\cite{Finn2017}, which learns a good initialization for fast adaptation.
    \item Ensemble of averages (EOA)~\cite{Arpit2022}, which trains an ensemble of independent moving average models.
    \item Song2022~\cite{Song2022}, an event-related desynchronization detection and false positive rejection algorithm based on the time-frequency characteristics of MI.
\end{enumerate}
Note that CEC and Song2022 were proposed for asynchronous MI classification, so their original algorithms were implemented. The other 9 algorithms cannot be directly used for asynchronous BCIs, so they were embedded into SWPC. More specifically, BYOL, SimCLR, MoCo, ContraWR, SSCL and Ou2022 were used to replace the SSL part of SWPC (the supervised learning part was kept), MAML and EOA were used to replace the supervised training part of SWPC (the SSL part was removed), and JTS was only used in supervised training of the prescreening module.

Both within-subject and cross-subject classifications were performed. For within-subject experiments, Session 1 was used for training, and Session 2 (with all triggers removed) of the same subject for testing. For cross-subject experiments, Session 2 of a subject was used for testing, and Session 1 from all other subjects were combined for training. 40\% of the training data were reserved as the validation set for determining the optimal hyperparameters, which were then applied to all training data to re-train the model. Except for CEC, which has no randomness, all other algorithms were repeated five times, and the average is reported.

\subsection{Results}  \label{subsec:result}

The ACCs and standard deviations (across different subjects) of different approaches on the four MI datasets in within-subject classification are shown in Tables~\ref{tab:MI1_within}-\ref{tab:MI4_within}, respectively. The ACCs and standard deviations in cross-subject classification are shown in Tables~\ref{tab:MI1_cross}-\ref{tab:MI4_cross}, respectively. The best results are marked in bold. Clearly, our proposed SWPC achieved the best average results on all four datasets in both within-subject and cross-subject classifications.

\begin{table*}[htpb] \centering \setlength{\tabcolsep}{3.2mm}
\caption{ACCs on MI1 in within-subject classification.} \label{tab:MI1_within}
\begin{tabular}{c|ccccccc|c}
\Xhline{1.2pt}
\multirow{2}{*}{Approach}
& \multicolumn{7}{c|}{Subject} & \multirow{2}{*}{Average} \\  \cline{2-8}
& 1     & 2     & 3     & 4     & 5     & 6     & 7     & \\  \hline
CEC
& 61.35 &51.57 &82.76  &52.75  &45.24 &85.45 &77.64 &65.25$\pm$16.47 \\
JTS
& 67.34& 51.88& 87.34& 55.62 & 50.24& 90.01& 82.57 &69.29$\pm$17.26\\	
BYOL
& 71.31 &52.56 &85.19 &53.94  &\textbf{54.64} &88.67 &77.56 &69.12$\pm$15.44 \\
SimCLR
& 67.06 &52.47 &85.81 &55.94 & 53.17& 88.58& 85.11 &69.73$\pm$16.43 \\
MoCo
& 65.28& 54.17& \textbf{87.51}& 56.25 & 54.17& 89.58& \textbf{86.81} &70.54$\pm$16.75 \\
ContraWR
& 68.45 &53.49 	&85.79 	&56.78 	&51.76 	&87.65 	&84.65 	&69.80$\pm$16.11  \\
SSCL
& 65.33 &51.43 	&83.88 	&53.45 	&52.21 	&84.67 	&81.17 	&67.45$\pm$15.51  \\
Ou2022
& 64.31 &52.34 	&84.79 	&54.67 	&51.56 	&86.65 	&83.45 	&68.25$\pm$16.20 \\
MAML
& 70.56 	&50.59 	&81.54 	&53.91 	&50.42 	&86.71 	&81.89  	&67.95$\pm$16.04  \\
EOA
& 71.21 	&52.81 	&83.81 	&57.91 	&51.09 	&\textbf{91.41} 	&79.61  	&69.69$\pm$16.03 \\
Song2022
& 71.57 	&\textbf{57.27} 	&85.31 	&55.12 	&51.89 	&86.51 	&85.21  	&70.41$\pm$15.55 \\\hline
SWPC
& \textbf{73.61}& 56.25& \textbf{87.51}& \textbf{58.33}& 52.78& 90.97& 84.72 &\textbf{72.02}$\pm$16.17 \\
\hline
\Xhline{1.2pt}
\end{tabular}
\end{table*}

\begin{table*}[htpb] \centering \setlength{\tabcolsep}{3.2mm}
\caption{ACCs on MI2 in within-subject classification.} \label{tab:MI2_within}
\begin{tabular}{c|ccccccc|c}
\Xhline{1.2pt}
\multirow{2}{*}{Approach}  & \multicolumn{7}{c|}{Subject} & \multirow{2}{*}{Average} \\  \cline{2-8}
& 1     & 2     & 3     & 4     & 5     & 6     & 7     &                      \\  \hline
CEC
& 37.12& 25.45& 48.23& 31.45 & 35.45& 52.14& 50.24 &40.01$\pm$10.28 \\
JTS
& 42.37& 27.45& 51.78& 37.69 & 38.31& 54.25& 51.27 &\multicolumn{1}{l}{43.30$\pm$9.69} \\
BYOL
& 51.12& 28.21& 59.11& 36.54 & 35.51& \textbf{61.19}& 58.07 &47.11$\pm$13.43 \\
SimCLR
& 52.86& 30.99& 58.76& 36.19 & 37.93& 58.76& 60.15 &47.95$\pm$12.47 \\
MoCo
& 55.99 &29.25 &55.99 &38.28 &36.89  &52.51  &61.19 &47.16$\pm$12.16 \\
ContraWR
& 54.78 &\textbf{33.45} 	&55.13 	&36.78 	&37.78 	&56.67 	&61.15 	&47.96$\pm$11.45 \\
SSCL
& 53.12 	&29.45 	&55.43 	&34.56 	&36.21 	&57.87 	&62.21 	&46.98$\pm$13.15  \\
Ou2022
& 52.86 	&28.56 	&55.12 	&31.34 	&35.12 	&54.12 	&62.12 	&45.61$\pm$13.49  \\
MAML
& 52.41 	&27.87 	&53.76 	&33.81 	&34.21 	&53.71 	&54.17  	&44.28$\pm$11.71 \\
EOA
& 55.79 	&31.98 	&55.93 	&35.76 	&38.12 	&37.21 	&61.62  	&45.20$\pm$12.07  \\
Song2022
& 53.54 	&30.46 	&57.21 	&38.28 	&35.27 	&55.41 	&60.31  	&47.21$\pm$12.12  \\ \hline
SWPC
& \textbf{57.64}& 31.25& \textbf{59.38}& \textbf{38.54}& \textbf{39.58}& 60.76& \textbf{63.89} &\textbf{50.15}$\pm$13.21 \\
\hline
\Xhline{1.2pt}
\end{tabular}
\end{table*}

\begin{table*}[htpb] \centering \setlength{\tabcolsep}{2mm}
\caption{ACCs on MI3 in within-subject classification.} \label{tab:MI3_within}
\begin{tabular}{c|ccccccccccccc|c}
\Xhline{1.2pt}
\multirow{2}{*}{Approach} & \multicolumn{13}{c|}{Subject} & \multirow{2}{*}{Average} \\  \cline{2-14}
& 1     & 2     & 3     & 4     & 5     & 6     & 7      & 8     & 9  & 10 & 11 & 12 & 13    &                      \\  \hline
CEC
& 51.1 &53.6 &45.2 &48.9 &47.1 &51.3 &58.1 &\textbf{56.9} &78.1 &57.8 &57.2 &\textbf{56.1} &43.1 &\multicolumn{1}{l}{54.2$\pm$8.7}\\
JTS
& 52.1 &57.8 &53.1 &57.8 &53.4 &54.2 &61.2 &56.3 &91.2 &52.1 &\textbf{63.6} &54.7 &52.3 &58.4$\pm$10.4\\
BYOL
& 54.0 &69.0 &65.6 &60.6 &60.6 &72.3 &69.0 &49.0 &99.0 &\textbf{64.0} &57.3 &54.0 &50.6 &63.5$\pm$12.9\\
SimCLR
& 51.3 &71.3& 64.7& 74.7& \textbf{64.7}& \textbf{76.3}& 64.7& 49.7& 93.0& 61.3& 49.7& 41.3& 53.0 &62.7$\pm$14.0 \\
MoCo
& 55.0 &66.7& \textbf{70.0}& 68.3& 55.0& 58.3& \textbf{73.3}& 50.0& 93.3& 55.0& 56.7& 50.0& 51.7 &61.8$\pm$12.3 \\
ContraWR
& 52.5 &64.1 &66.3 &72.2 &58.1 &66.2 &71.1 &51.3 &98.0 &58.3 &61.1 &53.7 &51.2 &63.4$\pm$12.6  \\
SSCL
& 51.2 &61.1	 &67.1 &70.2 &52.3 &64.5 &67.9 &52.1 &\textbf{99.1} &55.1 &59.1 &52.3 &48.5 &61.6$\pm$13.3 \\
Ou2022
& 49.2 &59.1	 &67.3 &71.1	 &54.3 &63.1	 &70.1 &51.3	 &98.0	 &56.2 &57.8	 &54.3 &53.2	 &61.9$\pm$12.9 \\
MAML
& 52.8 &62.9 &60.4 &74.2 &58.2 &70.3 &68.5 &49.4 	&98.0 &58.4 	&52.6 &49.5 	&51.5 &62.1$\pm$13.4  \\
EOA
& 52.7 &68.5 &61.2 &69.2 &57.3 &71.7 &65.3 &48.1 &98.2 &56.9 	&52.6 &47.8 &50.2  &61.5$\pm$13.7 \\
Song2022
& \textbf{57.8} &67.2 &63.9 &71.7 &55.5 &68.4 &64.3 &48.6 &98.0 &61.8 	&55.2 &50.2 	&52.1 &62.7$\pm$12.8 \\ \hline
SWPC
& 55.0 &\textbf{71.7}& 61.7& \textbf{76.7}& 60.0& 73.3& 71.7& 50.0& 98.3& 63.3& 56.7& 51.7& \textbf{55.0} &\textbf{65.0}$\pm$13.2 \\
\hline
\Xhline{1.2pt}
\end{tabular}
\end{table*}

\begin{table*}[htpb] \centering \setlength{\tabcolsep}{3.2mm}
\caption{ACCs on MI4 in within-subject classification.} \label{tab:MI4_within}
\begin{tabular}{c|ccccccc|c}
\Xhline{1.2pt}
\multirow{2}{*}{Approach} & \multicolumn{7}{c|}{Subject}& \multirow{2}{*}{Average} \\  \cline{2-8}
& 1     & 2     & 3     & 4     & 5     & 6     & 7            &                      \\  \hline
CEC
& 62.31& 89.23 &71.25 &72.15 &63.56 &76.23 & 76.56 &\multicolumn{1}{l}{73.04$\pm$9.07} \\
JTS
& 73.21& 94.34& 76.28& 75.45& 67.56 & 82.31& 85.43  &\multicolumn{1}{l}{79.23$\pm$8.87} \\
BYOL
& 75.62& 97.19& 78.44& 79.69& 77.81& 87.81& 86.88  &\multicolumn{1}{l}{83.35$\pm$7.66} \\
SimCLR
& 72.51& 97.52 & 71.88& 80.62& 77.81& 85.31& 86.25  &\multicolumn{1}{l}{81.70$\pm$8.96}\\
MoCo
& \textbf{77.81}& 96.88& 75.94& 83.44& 72.81& 88.12& 85.31 &\multicolumn{1}{l}{82.90$\pm$8.22}\\
ContraWR
& 72.12	&96.48	 &\textbf{81.41}	&78.34 &73.12 &83.12	 &83.65	&\multicolumn{1}{l}{81.18$\pm$8.16}  \\
SSCL
& 69.43	 &91.67	 &78.45	&73.43 &75.02 &78.43	 &81.43	&\multicolumn{1}{l}{78.27$\pm$7.09} \\
Ou2022
& 68.76	  &95.46 &78.34	 &75.87	&73.56 &76.12 &82.75	 &\multicolumn{1}{l}{78.69$\pm$8.54} \\
MAML
& 67.31 &93.71 &78.42 &80.61 &76.42 	&85.32 &84.02 &\multicolumn{1}{l}{80.83$\pm$8.21}  \\
EOA
& 72.76 &96.32  &75.64 &79.21 &77.37 &84.28 &85.21 &\multicolumn{1}{l}{81.54$\pm$7.90} \\
Song2022
& 75.82 	&97.61  &76.32 &76.32 &75.42 &85.65	&86.73 &\multicolumn{1}{l}{81.84$\pm$8.54} \\ \hline
SWPC
& 71.25 & \textbf{97.81} &80.01& \textbf{84.69}& \textbf{79.38}& \textbf{89.06}& \textbf{87.81}&\multicolumn{1}{l}{\textbf{84.29}$\pm$8.47} \\
\hline
\Xhline{1.2pt}
\end{tabular}
\end{table*}

\begin{table*}[htpb] \centering \setlength{\tabcolsep}{3.2mm}
\caption{ACCs on MI1 in cross-subject classification.} \label{tab:MI1_cross}
\begin{tabular}{c|ccccccc|c}
\Xhline{1.2pt}
\multirow{2}{*}{Approach} & \multicolumn{7}{c|}{Subject}                    & \multirow{2}{*}{Average} \\  \cline{2-8}
& 1     & 2     & 3   & 4     & 5     & 6     & 7          &                      \\  \hline
CEC
& 66.73& 45.12& 58.23& 53.23& 53.12& 72.13& 48.19 &\multicolumn{1}{l}{56.68$\pm$9.77} \\
JTS
& 75.43 & 48.21& 64.32& 61.78& 57.24& 79.25& 57.14 &\multicolumn{1}{l}{63.34$\pm$10.86} \\
BYOL
& 73.61& 51.39& 70.83& 60.42& 56.94& 85.42& 56.25 &\multicolumn{1}{l}{64.98$\pm$12.06} \\
SimCLR
& 77.78& 50.69& 72.92& 63.89& 54.86& 82.64& 49.31 &\multicolumn{1}{l}{64.58$\pm$13.48} \\
MoCo
& 81.94& 50.01& 59.03& \textbf{66.67}& 59.03& 85.42& 60.42 &\multicolumn{1}{l}{66.07$\pm$13.01} \\
ContraWR
& \textbf{82.54} &48.78 &66.54 &61.45 &59.13 &81.12 &57.78 &\multicolumn{1}{l}{65.33$\pm$12.46} \\
SSCL
&  74.34	 &49.45 &69.87 &58.78&58.87 &79.32 &54.17 &\multicolumn{1}{l}{63.54$\pm$11.08} \\
Ou2022
&  74.91 &48.76 &67.43 &56.91 &58.76 &75.54 &51.12 &\multicolumn{1}{l}{61.92$\pm$10.87} \\
MAML
& 76.51 &48.93 	&71.68 	&59.32 	&58.32 	&81.83 	&57.32   &\multicolumn{1}{l}{64.84$\pm$11.93} \\
EOA
&  77.81 &51.82 	&72.01 	&57.71 	&60.63 	&83.62 	&56.32   &\multicolumn{1}{l}{65.71$\pm$12.10} \\
Song2022
&  76.27 &50.86 	&74.32 	&60.91 	&57.36 	&83.91 	&57.84   &\multicolumn{1}{l}{65.92$\pm$12.19} \\ \hline
SWPC
& 80.56& \textbf{52.08}& \textbf{75.69}& 62.51 & \textbf{61.11}& \textbf{86.81}& \textbf{59.03} &\multicolumn{1}{l}{\textbf{68.26}$\pm$12.79} \\
\hline
\Xhline{1.2pt}
\end{tabular}
\end{table*}

\begin{table*}[htpb] \centering \setlength{\tabcolsep}{3.2mm}
\caption{ACCs on MI2 in cross-subject classification.} \label{tab:MI2_cross}
\begin{tabular}{c|ccccccc|c}
\Xhline{1.2pt}
\multirow{2}{*}{Approach} & \multicolumn{7}{c|}{Subject}   & \multirow{2}{*}{Average} \\  \cline{2-8}
& 1     & 2     & 3     & 4     & 5     & 6     & 7      &                      \\  \hline
CEC
& 42.17& 22.56& 36.75& 37.98& 32.87& 38.67& 41.25 &\multicolumn{1}{l}{36.04$\pm$6.68} \\
JTS
& 51.87& 25.37& 41.76& \textbf{45.67}& 35.43& 43.26& 45.23 &\multicolumn{1}{l}{41.23$\pm$8.55} \\
BYOL
& 52.47& 25.74& \textbf{50.04}& 39.62& 37.89& \textbf{52.12}& 42.41 &\multicolumn{1}{l}{42.90$\pm$9.64} \\
SimCLR
& 44.44& \textbf{29.17} & 45.83& 27.78& 28.82& 31.61& 32.29 &\multicolumn{1}{l}{34.28$\pm$7.59} \\
MoCo
& 41.67& 23.61& 43.41& 38.89& 34.03& 48.61& 39.24 &\multicolumn{1}{l}{38.49$\pm$7.95} \\
ContraWR
& 52.23	&25.54	&45.21	&38.32	&38.76	&45.87	&45.12	&\multicolumn{1}{l}{41.58$\pm$8.50}  \\
SSCL
& 49.87	&23.51	&43.93	&35.56 &38.78	&41.32	&43.87	&\multicolumn{1}{l}{39.55$\pm$8.38}  \\
Ou2022
& 48.76	&24.28	&43.87	&31.53	&36.65	&38.87	&42.46	&\multicolumn{1}{l}{38.06$\pm$8.19} \\
MAML
& 46.92 	&25.98 	&40.92 	&37.81 	&39.21 	&44.63 	&43.51 	&\multicolumn{1}{l}{39.85$\pm$6.88}  \\
EOA
& 53.85 &25.46 	&45.18 	&38.62 	&37.91 	&42.81 	&44.95 	&\multicolumn{1}{l}{41.25$\pm$8.73}  \\
Song2022
& 55.91 &24.74 	&46.32 	&40.83 	&38.62 	&48.65 	&46.94 	&\multicolumn{1}{l}{43.14$\pm$9.85} \\ \hline
SWPC
& \textbf{56.25}& 27.08& 48.96& 42.01 & \textbf{41.67}& 51.04& \textbf{48.26} &\multicolumn{1}{l}{\textbf{45.04}$\pm$9.40
} \\
\hline
\Xhline{1.2pt}
\end{tabular}
\end{table*}

\begin{table*}[htpb] \centering \setlength{\tabcolsep}{2mm}
\caption{ACCs on MI3 in cross-subject classification.} \label{tab:MI3_cross}
\begin{tabular}{c|ccccccccccccc|c}
\Xhline{1.2pt}
\multirow{2}{*}{Approach} & \multicolumn{13}{c|}{Subject}
& \multirow{2}{*}{Average} \\  \cline{2-14}
& 1     & 2     & 3     & 4     & 5     & 6     & 7      & 8     & 9  & 10 & 11 & 12 & 13    &                      \\  \hline
CEC
& \textbf{55.6}& 57.3& 62.1& 62.3& 51.3& 58.2& 57.4& 71.3& 74.3& 51.2& 42.1& 47.8& 51.2 &\multicolumn{1}{l}{57.1$\pm$9.1} \\
JTS
& 51.2 & 68.2& 64.3& 67.6& 58.2& 71.2& 63.5& 71.2& 75.5& 53.4& 45.7& 52.6& 50.2 &\multicolumn{1}{l}{61.2$\pm$9.6} \\
BYOL
& 44.7& 69.7& 69.7& 66.3& 66.3& 73.0& 68.0& 79.7& 84.7& 51.3& 44.7& 56.3& 48.0&\multicolumn{1}{l}{63.3$\pm$13.1} \\
SimCLR
& 46.7& 70.0& 68.3& \textbf{73.3}& 63.3& \textbf{76.7}& 58.3& 78.3& 85.0& \textbf{61.7}& 50.0& 55.0& 48.3 &\multicolumn{1}{l}{64.2$\pm$12.3} \\
MoCo
& 51.7& \textbf{76.7}& 66.7& \textbf{73.3}& 53.3& 65.0& 56.7& 65.0& \textbf{93.3}& 50.0& 50.0& \textbf{60.0}& 53.3& \multicolumn{1}{l}{62.7$\pm$12.7} \\
ContraWR
& 53.9 &70.3 &64.9 &71.1 &61.7 &75.4 &67.8 &76.5 &87.2 &52.8 &51.7 &52.3 &48.3 	&\multicolumn{1}{l}{64.1$\pm$11.9} \\
SSCL
& 52.3 	&67.1 &62.3 &67.3 &64.1 	&71.3 &65.7 	&71.3 &85.4 	&50.6 &52.1 &51.9 &\textbf{54.1} &\multicolumn{1}{l}{62.7$\pm$10.3} \\
Ou2022
& 53.1 	&65.4 &61.8 	&65.4 &67.2 	&67.8 &62.8 	&70.3 &81.5 &51.8 &54.4 &53.7 &52.8 &\multicolumn{1}{l}{62.2$\pm$8.8} \\
MAML
& 52.7 	&63.2 &65.9 &65.3 &63.6 &69.3 &65.3 &78.4 &85.3 &54.2 &48.2 &52.3 &48.3   &\multicolumn{1}{l}{62.5$\pm$11.3} \\
EOA
& 53.4 	&65.3 &68.2 &68.5 &65.4 	&73.2 &68.4 	&79.2 &82.5 	&52.3 &44.6 	&51.4 &50.4   &\multicolumn{1}{l}{63.3$\pm$11.8} \\
Song2022
& 52.3 	&68.4 &67.3 	&70.2 &61.4 	&72.9 &\textbf{72.5} &81.9 &83.7 &51.2 &\textbf{56.2} &50.2 &51.7   &\multicolumn{1}{l}{64.6$\pm$11.7} \\ \hline
SWPC
& 50.0& 73.3& \textbf{70.0}& 71.7& \textbf{68.3}& 75.0& 70.0& \textbf{83.3}& 88.3& 53.3& 46.7& 56.7& 51.7 &\multicolumn{1}{l}{\textbf{66.2}$\pm$13.2} \\
\hline
\Xhline{1.2pt}
\end{tabular}
\end{table*}

\begin{table*}[htpb] \centering \setlength{\tabcolsep}{3.2mm}
\caption{ACCs on MI4 in cross-subject classification.} \label{tab:MI4_cross}
\begin{tabular}{c|ccccccc|c}
\Xhline{1.2pt}
\multirow{2}{*}{Approach} & \multicolumn{7}{c|}{Subject}
& \multirow{2}{*}{Average} \\  \cline{2-8}
& 1     & 2     & 3     & 4     & 5     & 6     & 7          &      \\  \hline
CEC
& 60.15& 81.23& 75.45& 72.34& 68.43& 76.54& 78.23 &\multicolumn{1}{l}{73.20$\pm$7.07} \\
JTS
& 67.32& 86.24& 75.34& 78.12& 75.23& \textbf{86.12}& 81.23 &\multicolumn{1}{l}{78.51$\pm$6.72} \\
BYOL
&\textbf{77.81}& 87.19& 72.51& \textbf{82.81}& 64.06& 84.69& 72.50 &\multicolumn{1}{l}{77.37$\pm$8.21} \\
SimCLR
& 62.81& 90.62& 67.19& 62.81& 69.06& 78.44& 78.75&\multicolumn{1}{l}{72.81$\pm$10.25} \\
MoCo
& 73.75& 90.62& 74.38& 80.62& \textbf{76.25}& 79.69& 50.62&\multicolumn{1}{l}{75.13$\pm$12.22} \\
ContraWR
& 72.45  &84.98 	&74.98 	&78.42 	&72.98 	&82.76 	&81.65 	&\multicolumn{1}{l}{78.32$\pm$4.99} \\
SSCL
& 67.53  &85.43 	&71.87 	&74.91 	&67.43 	&81.44 	&77.65 	&\multicolumn{1}{l}{75.18$\pm$6.83} \\
Ou2022
& 66.37	&81.91 	&73.42 	&80.41 	&71.25 	&80.41 	&73.62 	&\multicolumn{1}{l}{75.34$\pm$5.75} \\
MAML
& 68.32 	&85.43 	&72.91 	&73.98 	&73.81 	&72.96 	&83.51  	&\multicolumn{1}{l}{75.85$\pm$6.22} \\
EOA
& 72.51 	&83.81 	&75.73 	&80.61 	&69.61 	&78.71 	&82.61  	&\multicolumn{1}{l}{77.66$\pm$5.28} \\
Song2022
& 76.26 	&87.53 	&73.61 	&82.71 	&73.27 	&73.81 	&80.51  	&\multicolumn{1}{l}{78.24$\pm$5.49} \\ \hline
SWPC
& 70.01 & \textbf{91.88}& \textbf{76.88}& 80.94& 75.94& 80.62& \textbf{85.01}&\multicolumn{1}{l}{\textbf{80.18}$\pm$6.99} \\
\hline \Xhline{1.2pt}
\end{tabular}
\end{table*}

We also studied the effectiveness of SSL to the prescreening module and the classification module. Table~\ref{tab:TPR_presceening} shows the MI identification accuracies of the prescreening module, with and without SSL. Table~\ref{tab:ACC_classification} shows the ACCs of the classification module, assuming there are triggers. Clearly, SSL was always beneficial to both the prescreening module and the classification module.

\begin{table*}[htpb]
\centering \setlength{\tabcolsep}{4mm}
\renewcommand{\arraystretch}{1.1}
\caption{MI identification accuracies of the prescreening module.} \label{tab:TPR_presceening}
\begin{tabular}{c|c|cccc|c}
\Xhline{1.2pt}
\multicolumn{2}{c|}{Scenario} & MI1 & MI2  & MI3  & MI4   & Average            \\ \hline
\multirow{2}{*}{\begin{tabular}[c]{@{}c@{}}Cross-subject\end{tabular}} & \multirow{1}{*}{Without SSL in Prescreening}
& 90.74 & 85.80 & 89.14 & 94.62 & 90.08         \\ \cline{2-7}
& With SSL in Prescreening     & \textbf{92.69} & \textbf{86.92} & \textbf{91.89} & \textbf{94.17} & \textbf{91.41} \\ \Xhline{1.2pt}
\multirow{2}{*}{\begin{tabular}[c]{@{}c@{}}Within-subject\end{tabular}} & \multirow{1}{*}{Without SSL in Prescreening}
& 91.52 & 87.42 & 91.26 & 95.06 & 91.35          \\ \cline{2-7}
& With SSL in Prescreening    & \textbf{93.74} & \textbf{89.19} & \textbf{93.21} & \textbf{96.19} & \textbf{93.08} \\ \Xhline{1.2pt}
\end{tabular}
\end{table*}

\begin{table*}[htpb]
\centering \setlength{\tabcolsep}{4mm}
\renewcommand{\arraystretch}{1.1}
\caption{ACCs of the classification module, when there are triggers.} \label{tab:ACC_classification}
\begin{tabular}{c|c|cccc|c}
\Xhline{1.2pt}
\multicolumn{2}{c|}{Scenario} & MI1 & MI2  & MI3  & MI4   & Average            \\ \hline
\multirow{2}{*}{\begin{tabular}[c]{@{}c@{}}Cross-subject\end{tabular}} & \multirow{1}{*}{Without SSL in Classification}
& 72.21 & 50.41 & 72.07 & 82.77 &69.37          \\ \cline{2-7}
& With SSL in Classification     & \textbf{75.39} & \textbf{52.95} & \textbf{73.81} & \textbf{84.56} & \textbf{71.67} \\ \Xhline{1.2pt}
\multirow{2}{*}{\begin{tabular}[c]{@{}c@{}}Within-subject\end{tabular}} & \multirow{1}{*}{Without SSL in Classification}
& 77.56 & 55.53 & 70.48 & 86.51 &72.52           \\ \cline{2-7}
& With SSL in Classification    & \textbf{79.83} & \textbf{56.79} & \textbf{72.14} & \textbf{88.34} & \textbf{74.28} \\ \Xhline{1.2pt}
\end{tabular}
\end{table*}

Paired $t$-tests were performed to evaluate whether the performance improvements of SWPC over others were statistically significant in within-subject and cross-subject classifications. The results are shown in Tables~\ref{tab:ttest_within} and \ref{tab:ttest_cross}, respectively, where $p$-values smaller than 0.05 are marked with asterisks. Most of the performance improvements were statistically significant; particularly, in cross-subject classification, SWPC statistically significantly outperformed each algorithm on at least three datasets.

\begin{table}[h] \centering \setlength{\tabcolsep}{3mm} \centering
\renewcommand{\arraystretch}{1.1}
\caption{Adjusted $p$-values of paired $t$-tests between SWPC and other approaches in within-subject classification.} \label{tab:ttest_within}
\begin{tabular}{c|llll}\Xhline{1.2pt}
SWPC vs. &\multicolumn{1}{c}{MI1} &\multicolumn{1}{c}{MI2} &\multicolumn{1}{c}{MI3} &\multicolumn{1}{c}{MI4} \\
\hline
CEC       &0.0140$^{\ast}$  &0.0018$^{\ast}$  & 0.0172$^{\ast}$  & 0.0041$^{\ast}$   \\
JST       &0.0140$^{\ast}$  &0.0015$^{\ast}$  & 0.0126$^{\ast}$  & 0.0140$^{\ast}$    \\
BYOL      &0.0538           &0.0721           & 0.0140$^{\ast}$  & 0.0467$^{\ast}$    \\
SimCLR    &0.0647           &0.0361$^{\ast}$  & 0.0752           & 0.0132$^{\ast}$    \\
MoCo      &0.0291$^{\ast}$  &0.0227$^{\ast}$  & 0.0698           & 0.0231$^{\ast}$    \\
ContraWR  &0.0783           &0.0658           & 0.0140$^{\ast}$  & 0.0729  \\
SSCL    &0.0140$^{\ast}$  &0.0140$^{\ast}$  & 0.0545           & 0.0140$^{\ast}$  \\
Ou2022 &0.0140$^{\ast}$  &0.0140$^{\ast}$  & 0.0140$^{\ast}$  & 0.0158$^{\ast}$ \\
MAML      &0.0778           &0.0227$^{\ast}$  & 0.0782           & 0.0136$^{\ast}$  \\
EOA       &0.0432$^{\ast}$  &0.0285$^{\ast}$  & 0.0589           & 0.0267$^{\ast}$  \\
Song2022      &0.0231$^{\ast}$  &0.0782           & 0.0140$^{\ast}$  & 0.0159$^{\ast}$ \\
\Xhline{1.2pt}
\end{tabular}
\end{table}%

\begin{table}[h] \centering \setlength{\tabcolsep}{3mm} \centering
\renewcommand{\arraystretch}{1.1}
\caption{Adjusted $p$-values of paired $t$-tests between SWPC and other approaches in cross-subject classification.} \label{tab:ttest_cross}
\begin{tabular}{c|llll}\Xhline{1.2pt}
SWPC vs. &\multicolumn{1}{c}{MI1} &\multicolumn{1}{c}{MI2} &\multicolumn{1}{c}{MI3} &\multicolumn{1}{c}{MI4} \\
\hline
CEC       &0.0037$^{\ast}$  &0.0067$^{\ast}$  & 0.0013$^{\ast}$  & 0.0027$^{\ast}$   \\
JST       &0.0167$^{\ast}$  &0.0196$^{\ast}$  & 0.0085$^{\ast}$  & 0.0956             \\
BYOL      &0.0187$^{\ast}$  &0.0063$^{\ast}$  & 0.0018$^{\ast}$  & 0.0073$^{\ast}$    \\
SimCLR    &0.0213$^{\ast}$  &0.0196$^{\ast}$  & 0.1088           & 0.0027$^{\ast}$    \\
MoCo      &0.0752           &0.0027$^{\ast}$  & 0.0196$^{\ast}$  & 0.0031$^{\ast}$    \\
ContraWR  &0.0292$^{\ast}$  &0.0596           & 0.0178$^{\ast}$  & 0.0386$^{\ast}$    \\
SSCL    &0.0358$^{\ast}$  &0.0196$^{\ast}$  & 0.0478$^{\ast}$  & 0.0196$^{\ast}$    \\
Ou2022 &0.0169$^{\ast}$  &0.0196$^{\ast}$  & 0.0523           & 0.0196$^{\ast}$    \\
MAML      &0.0782           &0.0596           & 0.0178$^{\ast}$  & 0.0386$^{\ast}$    \\
EOA       &0.0358$^{\ast}$  &0.0196$^{\ast}$  & 0.0478$^{\ast}$  & 0.0196$^{\ast}$    \\
Song2022      &0.0171$^{\ast}$  &0.0782           & 0.0523           & 0.0782             \\
\Xhline{1.2pt}
\end{tabular}
\end{table}%

\subsection{Ablation Study}

Ablation studies were performed to evaluate if SSL in the prescreening module and the classification module, and the final averaging, are truly necessary and beneficial. Within-subject and cross-subject classification results are shown in Tables~\ref{tab:ablation within} and \ref{tab:ablation cross}, respectively. Clearly, all three components were essential to the superior performance of SWPC.

\begin{table*}[htpb] \centering \setlength{\tabcolsep}{4mm}
\renewcommand{\arraystretch}{1.0}
\caption{Ablation study results in within-subject classification.} \label{tab:ablation within}
\begin{tabular}{ccc|cccc|c} \Xhline{1.2pt}
SSL in Prescreening& SSL in Classification & Averaging& MI1   & MI2   & MI3   & MI4   & Average  \\ \hline
×         & ×        & ×        & 68.36 & 44.01 & 58.52 & 81.32 & 63.05  \\ \hline
$\surd$   & ×        & ×        & 69.45 & 46.12 & 60.04 &	82.15 & 64.44  \\
×         & $\surd$  & ×        & 70.93 & 48.43 & 64.03 &	82.34 & 66.43  \\
×         & ×        & $\surd$  & 69.76 & 47.17 &	 59.13 &	83.43 &	64.87   \\ \hline
$\surd$   & $\surd$  & ×        & 71.24 & 49.31 & 64.22 &	83.39 &	67.04  \\
$\surd$   & ×        & $\surd$  & 70.98 &	 48.54 & 64.32 &	82.24 &	66.52 \\
×         & $\surd$  & $\surd$  & 71.87 & 49.41 &	 64.15 &	83.91 &	67.34  \\ \hline
$\surd$   & $\surd$  & $\surd$  & \textbf{72.02} & \textbf{50.15} & \textbf{65.00} & \textbf{84.29} & \textbf{67.87} \\ \Xhline{1.2pt}
\end{tabular}
\end{table*}

\begin{table*}[htpb] \centering \setlength{\tabcolsep}{4mm}
\renewcommand{\arraystretch}{1.0}
\caption{Ablation study results in cross-subject classification.} \label{tab:ablation cross}
\begin{tabular}{ccc|cccc|c} \Xhline{1.2pt}
SSL in Prescreening& SSL in Classification & Averaging& MI1   & MI2   & MI3   & MI4   & Average  \\ \hline
×         & ×        & ×        & 63.52 & 41.52 & 62.04 & 78.56 & 61.41  \\ \hline
$\surd$   & ×        & ×        & 65.65 & 42.53 & 64.36 & 79.32 & 62.97 \\
×         & $\surd$  & ×        & 66.81 & 43.47 & 65.17 & 79.54 & 63.75 \\
×         & ×        & $\surd$  & 64.25 & 42.18 & 63.76 & 78.47 & 62.17 \\ \hline
$\surd$   & $\surd$  & ×        & 67.16 & 44.27 & 65.32 & 79.57 & 64.08 \\
$\surd$   & ×        & $\surd$  & 65.51 & 43.25 & 64.24 & 78.32 & 62.83 \\
×         & $\surd$  & $\surd$  & 67.37 & 44.24 & 64.56 & 79.56 & 63.93  \\ \hline
$\surd$   & $\surd$  & $\surd$  & \textbf{68.26} & \textbf{45.04} & \textbf{66.20} & \textbf{80.18} & \textbf{64.88} \\ \Xhline{1.2pt}
\end{tabular}
\end{table*}

\subsection{Parameter Sensitivity Analysis} \label{subsec:parameter sensitivity analysis}

\begin{figure}[htbp]\centering
\subfigure[]{\label{fig:ACC_tau_within}    \includegraphics[width=0.7\linewidth,clip]{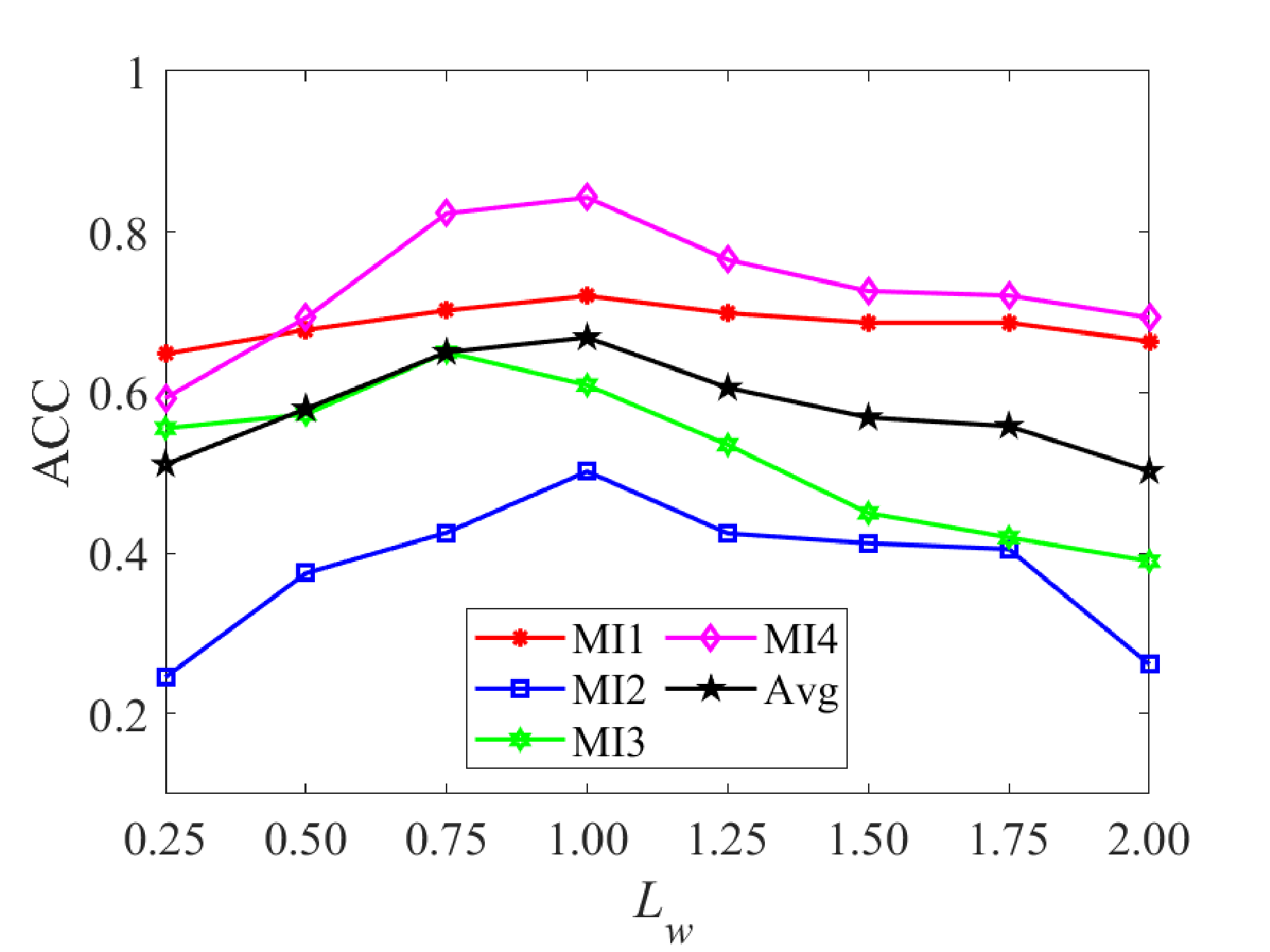}}
\subfigure[]{\label{fig:ACC_tau_cross}    \includegraphics[width=0.7\linewidth,clip]{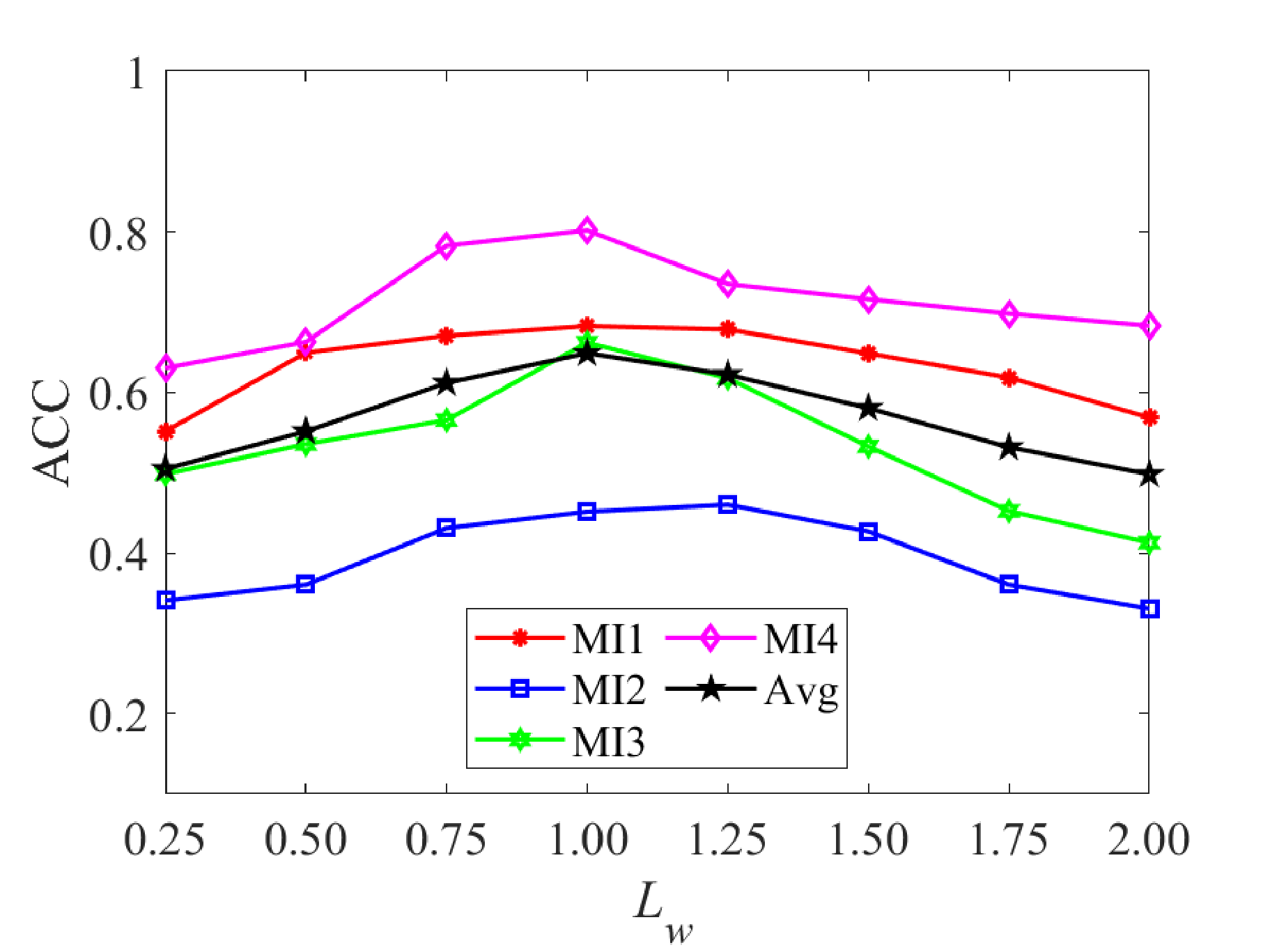}}
\caption{Change of the classification accuracy w.r.t. the time window length $L_w$ in (a) within-subject classification; and, (b) cross-subject classification.} \label{fig:sen}
\end{figure}

\begin{figure}[htbp]\centering
\subfigure[]{\label{fig:ACC_tau_within}    \includegraphics[width=0.7\linewidth,clip]{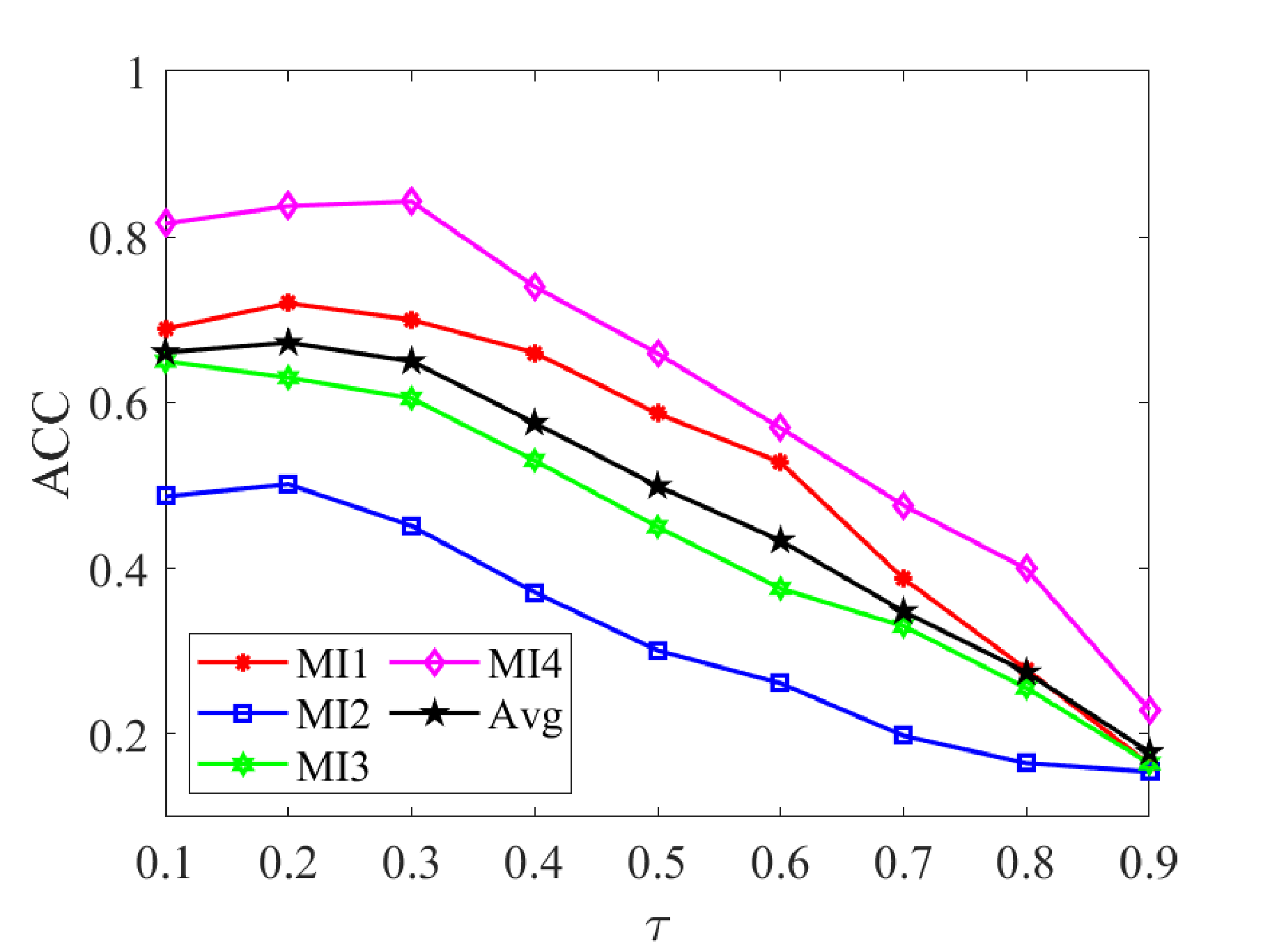}}
\subfigure[]{\label{fig:ACC_tau_cross}    \includegraphics[width=0.7\linewidth,clip]{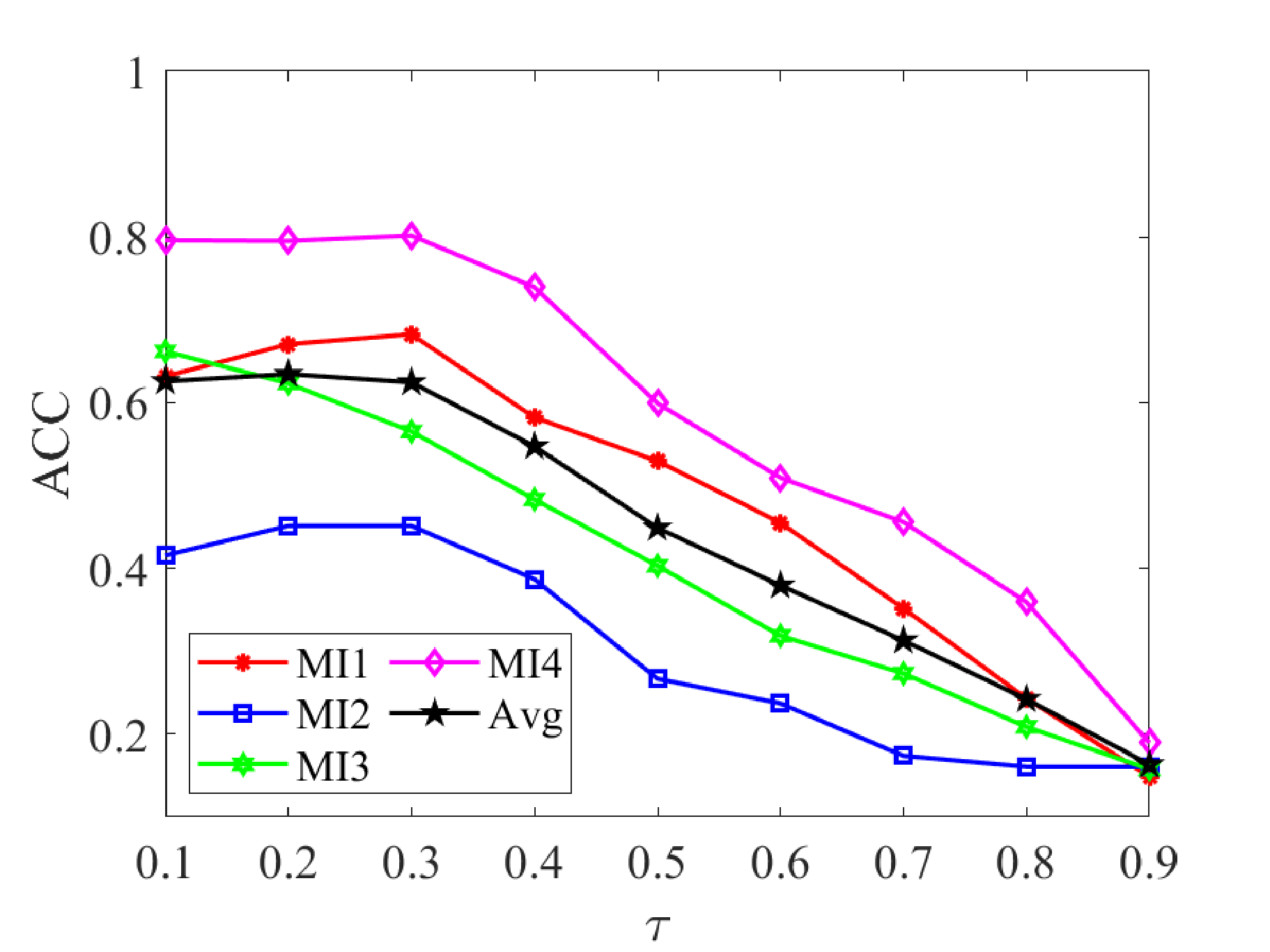}}
\caption{Change of the classification accuracy w.r.t. the prescreening threshold $\tau$ in (a) within-subject classification; and, (b) cross-subject classification.} \label{fig:tau}
\end{figure}

This subsection evaluates the sensitivity of SWPC to the time window length $L_w$ and the prescreening threshold $\tau$. The results are shown in Figs.~\ref{fig:sen} and \ref{fig:tau}, respectively. $L_w=1$ and $\tau=0.2$ seem to achieve the overall best performance on all datasets.

\subsection{Offline Classification}  \label{subsec:SWPC with access to test data}

All previous subsections considered online classification, i.e., the test data are available on-the-fly. This subsection further considers offline classification, where all test EEG data are available.

In offline classification, SSL on the test set $\mathcal{D}_t=\{ X_i^t \}^{n_t}_{i=1}$ (instead of on the training set $\mathcal{D}_s$ in online classification) may be used to improve the performance. Specifically, we first conducted SSL in Section~\ref{subsec:SSL for IM} on $\mathcal{D}_t$, and then SSL in Section~\ref{subsec:SSL for CM} on $\mathcal{\hat{D}}_t$, which consisted of EEG trials predicted as MI in $\mathcal{D}_t$. Tables~\ref{tab:with access to test data within} and \ref{tab:with access to test data cross} show the average results on the four datasets in within-subject and cross-subject classification, respectively. Offline classification accuracies were higher than their online counterparts for all approaches in both scenarios, because SSL on the test data themselves extracted more tailored features.

\begin{table}[t] \centering \setlength{\tabcolsep}{5mm}
\renewcommand{\arraystretch}{1.0}
\caption{Online and offline within-subject classification accuracies.} \label{tab:with access to test data within}
\begin{tabular}{c|c|c}
\Xhline{1.2pt}
\multirow{2}{*}{Approach} & \multicolumn{2}{c}{Scenario}
 \\  \cline{2-3}
& Online & Offline   \\  \hline
BYOL       & 65.76  & 66.63\\
SimCLR     & 65.53  & 67.83 \\
MoCo       & 65.60  & 66.54 \\
ContraWR   & 65.58  & 67.83\\
SSCL     & 63.57  & 64.87  \\
Ou2022  & 63.62  & 65.72\\ \hline
SWPC       & \textbf{67.87} & \textbf{69.10}\\
\Xhline{1.2pt}
\end{tabular}
\end{table}

\begin{table}[t] \centering \setlength{\tabcolsep}{5mm}
\renewcommand{\arraystretch}{1.0}
\caption{Online and offline cross-subject classification accuracies.} \label{tab:with access to test data cross}
\begin{tabular}{c|c|c}
\Xhline{1.2pt}
\multirow{2}{*}{Approach} & \multicolumn{2}{c}{Scenario}
 \\  \cline{2-3}
& Online & Offline   \\  \hline
BYOL       & 62.13  & 64.64\\
SimCLR     & 58.97  & 62.83 \\
MoCo       & 60.60  & 63.54 \\
ContraWR   & 62.34  & 65.43\\
SSCL     & 60.25  & 62.76  \\
Ou2022  & 59.37  & 63.21\\ \hline
SWPC       & \textbf{64.88} & \textbf{67.82}\\
\Xhline{1.2pt}
\end{tabular}
\end{table}

\section{Conclusions and Future Research} \label{sec:conclusion}

Asynchronous MI-based BCIs aim to detect the user's MI without explicit triggers. They are challenging to implement, because the algorithm needs to first distinguish between resting-states and MI trials, and then classify the MI trials into the correct task, all without any triggers. This paper has proposed SWPC for MI-based asynchronous BCIs, which consists of two modules: a prescreening module to screen MI trials from the resting-state, and a classification module for MI classification. Both modules are trained with supervised learning followed by SSL. Within-subject and cross-subject asynchronous MI classification on four different EEG datasets validated the effectiveness of SWPC, particularly, SSL to refine the feature extractors.

Our future research directions include:
\begin{enumerate}
\item \emph{Transfer learning}: Transfer learning can further mitigate cross-subject and cross-session data discrepancies. For asynchronous BCIs, data alignment~\cite{He2019}, source-free domain adaptation~\cite{Xia2022}, and domain generalization~\cite{Li2022} approaches may be used to further improve performance and protect user privacy.

\item \emph{Test-time adaptation}: Test-time adaptation updates the classifier using online unlabeled data to improve its performance. Our recent work~\cite{Li2023} has demonstrated its promising performance in synchronous MI-based BCIs, but how to apply it to asynchronous BCIs requires further investigation.

\item \emph{More BCI paradigms}: Only MI was considered in this paper. It is interesting to study if SWPC can be extended to other classical BCI paradigms, e.g., event-related potential and steady-state visual evoked potential.
\end{enumerate}


\end{document}